\documentclass[sigconf]{acmart}

\AtBeginDocument{%
  \providecommand\BibTeX{{%
    \normalfont B\kern-0.5em{\scshape i\kern-0.25em b}\kern-0.8em\TeX}}}

\copyrightyear{2021}
\setcopyright{iw3c2w3}
\acmYear{2021}
\acmConference[WWW '21]{Proceedings of the Web Conference 2021}{April 19--23, 2021}{Ljubljana, Slovenia}
\acmBooktitle{Proceedings of the Web Conference 2021 (WWW '21), April 19--23, 2021,
Ljubljana, Slovenia}

\acmPrice{}
\acmDOI{}
\acmISBN{978-1-4503-8312-7/21/04}

\pdfoutput=1
\usepackage{enumitem}
\usepackage{endnotes}
\usepackage{wrapfig}
\usepackage{color}
\usepackage{booktabs}
\usepackage{listings}
\usepackage{arydshln}
\usepackage{array}
\usepackage{tabularx}
\usepackage{makecell}
\usepackage{xspace}
\usepackage{balance}
\usepackage{multirow}
\usepackage{bigstrut}
\usepackage[super]{nth}
\usepackage{tikz}
\usepackage{stackengine}
\usepackage{hyperref}
\usepackage{todonotes}
\usepackage[utf8]{inputenc}

\usepackage{subfigure}
\usepackage{epsfig}

\DeclareUnicodeCharacter{394}{\change}

\newcommand{\point}[1]{\vspace{.05in} \par\noindent\textbf{#1}. }

\definecolor{brickred}{rgb}{0.8, 0.25, 0.33}

\newif\ifcomment
\commenttrue

\ifcomment
    \newcounter{MVNumberOfComments}
    \stepcounter{MVNumberOfComments}
    \newcommand{\mvnote}[1]{\textcolor{blue}{\small \bf [MV\#\arabic{MVNumberOfComments}\stepcounter{MVNumberOfComments}: #1]}}   
    
    \newcounter{CKNumberOfComments}
    \stepcounter{CKNumberOfComments}
    \newcommand{\conor}[1]{\textcolor{orange}{\small \bf [CK\#\arabic{CKNumberOfComments}\stepcounter{CKNumberOfComments}: #1]}}
    
    \newcounter{AANumberOfComments}
    \stepcounter{AANumberOfComments}
    \newcommand{\aanote}[1]{\textcolor{brickred}{\small \bf [AA\#\arabic{AANumberOfComments}\stepcounter{AANumberOfComments}: #1]}}   
\else
    \newcommand\mvnote[1]{}    
    \newcommand\conor[1]{}
    \newcommand\aanote[1]{}
    
\fi

\newcommand{\eg}{e.g., }
\newcommand{\ie}{i.e., }

\newcommand{\ourwork}{\textsc{BrowseLite}\xspace}
\newcommand{\rewriting}{\textsc{URL Rewriting}\xspace}
\newcommand{\fetching}{\textsc{image fetch reduction}\xspace}
\newcommand{\webcompat}{Web-compat\xspace}
\newcommand{\topHun}{top100\xspace}
\newcommand{\aprFty}{apr50k\xspace}
\newcommand{\aprHun}{apr100k\xspace}

\begin{document}


\title{\ourwork: A Private Data Saving Solution for the Web}


\author{Conor Kelton}
\affiliation{%
 \institution{Stony Brook University}
 \country{}}
 
\author{Matteo Varvello}
\affiliation{%
 \institution{Nokia Bell Labs}
 \country{}
}

\author{Andrius Aucinas}
\affiliation{%
 \institution{Brave Software}
 \country{}}

\author{Benjamin Livshits}
\affiliation{%
 \institution{Brave Software}
 \country{}}
\affiliation{
    \institution{Imperial College London}
    \country{}
}

\renewcommand{\shortauthors}{Kelton et al.}

\begin{abstract}
The median webpage has increased in size by more than~80\% in the last~4 years. This extra complexity allows for a rich browsing experience, but it hurts the majority of mobile users which still pay for their traffic. This has motivated several data-saving solutions, which aim at reducing the complexity of webpages by transforming their content. Despite each method being unique, they either reduce user privacy by further centralizing web traffic through data-saving middleboxes or introduce web compatibility (\webcompat) issues by removing content that breaks pages in unpredictable ways.

In this paper, we argue that data-saving is still possible without impacting either users privacy or \webcompat. Our main observation is that Web \textit{images} make up a large portion of Web traffic and have negligible impact on \webcompat. To this end we make two main contributions. First, we quantify the potential savings that image manipulation, such as dimension resizing, quality compression, and transcoding, enables at large scale:~300 landing and~880 internal pages. Next, we design and build \ourwork, an entirely client-side tool that achieves such data savings through opportunistically instrumenting existing server-side tooling to perform image compression, while simultaneously reducing the total amount of image data fetched. The effect of \ourwork on the user experience is quantified using standard page load metrics and a real user study of over~200 users across~50 optimized web pages. \ourwork allows for similar savings to middlebox approaches, while offering additional security, privacy, and \webcompat guarantees.
\end{abstract}

\ccsdesc[500]{Networks~Network performance analysis}
\ccsdesc[500]{Networks~Network experimentation}

\keywords{Web, Browsers, Performance, Optimizations, Images, Data saving}


\maketitle

\section{Introduction}
\label{sec:intro}
A multitude of studies from academia and industry alike suggest up-trends in the complexity and size of the mobile Web~\cite{Agababov2015-flywheel, http-archive-almanac, klotski-nsdi-2015, ChromeLiteMode, li-nsdi-2016}, so much so, that the median page now has reached ~2~MB, up over~80\% from~2016~\cite{http-archive-almanac}. While this complexity has undoubtedly created a richer browsing experience, it does cause downsides for mobile users. Byte-heavy pages are responsible for frustratingly slow browsing experiences on slower networks, along with significant monetary costs for users on limited mobile data plans. In Canada, for example, the median webpage costs~\$0.24 to load~\cite{what-does-my-site-cost}. 

As a result, emphasis has been placed on changing Web browsing to consume less data~\cite{Agababov2015-flywheel, ChromeLiteMode, opera-turbo, Netravali2019-prophecy, Singh2015-flexiweb}. While there has been some public confusion over the exact workings and reach of such \textit{data-saving} methods~\cite{tim-kadlec-making-sense}, recent studies of their actual implementations~\cite{simon-data-saving, Kondracki-2020-meddling, Tahir2020-deconstructing} reveal a few key shortcomings. 

First and foremost, these solutions impose various privacy concerns to their users when compared to regular Web browsing. Some are deployed as \textit{middlebox} services which either transparently proxy the user's unencrypted traffic~\cite{Agababov2015-flywheel, Singh2015-flexiweb}, or, apply URL redirection~\cite{GoogleWebLight} or Man-in-the-Middle proxies~\cite{wang-shandian-2016, Netravali2019-prophecy, opera-turbo} to also operate on encrypted traffic (HTTPS)~\cite{Kondracki-2020-meddling, Tahir2020-deconstructing}. Given the rise of HTTPS~\cite{felt-https-2016}, the former sees limited use, while the latter breaks the end-to-end principles of TLS, exposing potentially private or personalized Web contents to third parties~\cite{Kondracki-2020-meddling, Tahir2020-deconstructing}.

Further, the exact measures these systems take to actually save data are often cryptic~\cite{Kondracki-2020-meddling, simon-data-saving, Tahir2020-deconstructing, tim-kadlec-making-sense}. As determining \webcompat issues, or the ability to quantify broken webpages, is an open problem that requires large amount of manual effort~\cite{mozilla-webcompat}, it is hard to determine how and when these solutions actually break webpages; though when they do so there is usually public outcry~\cite{google-lite-break-pages}.

The goal of this work is to devise a data-saving solution which solves the above shortcomings. Our intuition is twofold: first, such a solution needs to be \textit{client-side} in order to eliminate the privacy and reach concerns of middlebox approaches. Namely, a client-side only approach has the potential to save data for personalized webpage content without exposing it to third parties. Second, by being \textit{image-centric} it can impose virtually no impact on \webcompat, in comparison to, for example, JavaScript code elision~\cite{Netravali2019-prophecy, GoogleWebLight, Chaqfeh-2020-cleaner}. As the median webpage is comprised of~900~KB of images (or about 44\% of its total size)~\cite{http-archive-almanac}, an image-based solution still has high potential for data-savings. 

Many of the aforementioned middlebox techniques~\cite{Agababov2015-flywheel, Singh2015-flexiweb} automatically apply popular image manipulation techniques (resizing, quality compression, and transcoding) to save data before the last mile (see Section~\ref{sec:related}). While the weight of images across webpages is known to be quite high, it is unclear which fraction of pages in the wild can benefit from such techniques. Our first contribution is to quantify the impact such middlebox techniques have when optimizing images. This quantification represents an upper bound of the data savings which a more private client-side solution can potentially obtain. We analyze such techniques by compressing webpages across a crawl of size~$\sim$1.2k (300 landing and 880 internal webpages), in which we find~21\% of total page weight can be saved at the median, and up to~90\% at the~95th percentile (depending on webpage rank and presence of cold or hot caches).

Motivated by this availability for image savings, we propose \ourwork to optimize images in a two-fold, more private, fashion. First, \ourwork takes advantage of existing compression technologies typically found on Web servers~\cite{image-cdns}, instrumenting them from the client at run-time. \ourwork uses information from images in the Document Object Model~(DOM) to force their compression while balancing their visual quality. This component of \ourwork, \rewriting, identifies images as candidates for compression using simple URL rewriting rules, which enable~70\% reduction of image sizes on~$\sim$16\% of webpages from our crawls.

From there, \ourwork\ takes a second approach to save-data for more general pages. This approach, which we call \fetching, uses a component of the HTTP standard known as \textit{range requests} to actually fetch less data from all Web images, thereby reducing their network data usage. To alleviate the impact on the user's quality of experience (QoE) caused by rendering only the requested portions of images, \ourwork differentiates between two standard image types on the Web, \textit{baseline} and \textit{progressive} images, during the page load. We show progressive images can be rendered almost completely with only a fraction of their data requested. For baseline images, we introduce our own technique, reflection, to make empty spaces on the webpage from these partial images appear less visually broken. We also show such techniques incur a modest trade-off in user QoE using both systematic metrics~\cite{speed-index} and real user studies across 50 of our pages with 200 crowdsourced users (See Section~\ref{sec:results}).

Our experiments show that \ourwork improves the bandwidth consumption of pages, taken from the same $\sim$1.2k webpages of our crawls, by~25\% at the median. While the primary function of \ourwork is to save data, our experiments on the pages of our crawls loaded over two network conditions show little effect, or even slight improvements, to the SpeedIndex~\cite{speed-index} of 80\% of pages, while causing modest overheads ($<$500ms) for the latter 20\%. Further, our comparisons of \ourwork to existing middlebox image reduction approaches show that savings from \ourwork (Section~\ref{sec:results}) outreach those offered via plain HTTP proxies, and are competitive with MITM middleboxes. While our comparisons to a real data-saving system which optimizes full contents of webpages, Google Web Light~\cite{GoogleWebLight}, do show the high potential of such tools to save data, 
we believe \ourwork to be a competitive alternative given its higher privacy, security, and \webcompat guarantees.
\section{Background and Related Work}
\label{sec:related}

\begin{figure*}[thb]
   \centering
   \subfigure[]{\psfig{figure=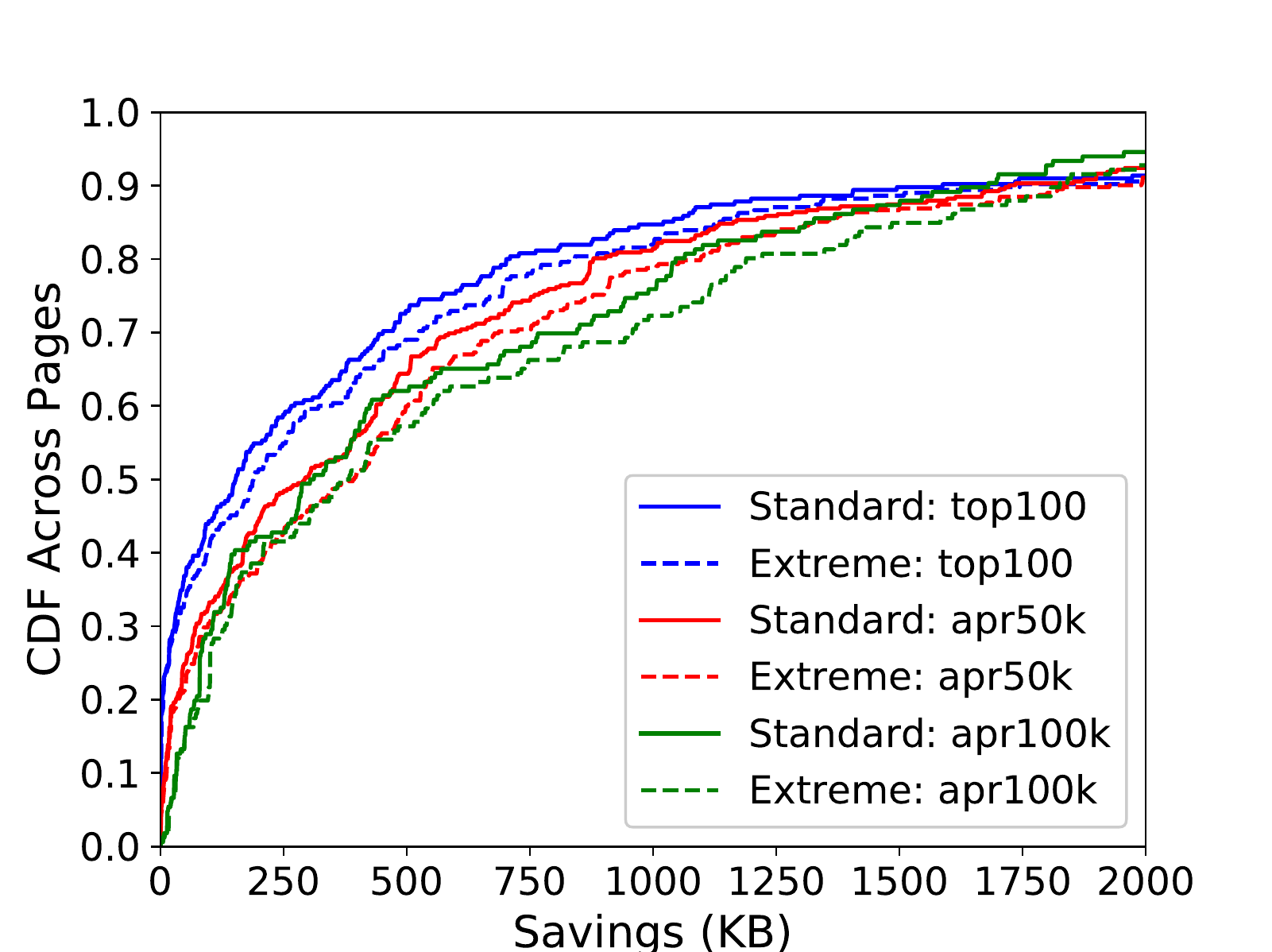, width=2.2in}\label{fig:raw-savings-and-ssim-a}}
   \subfigure[]{\psfig{figure=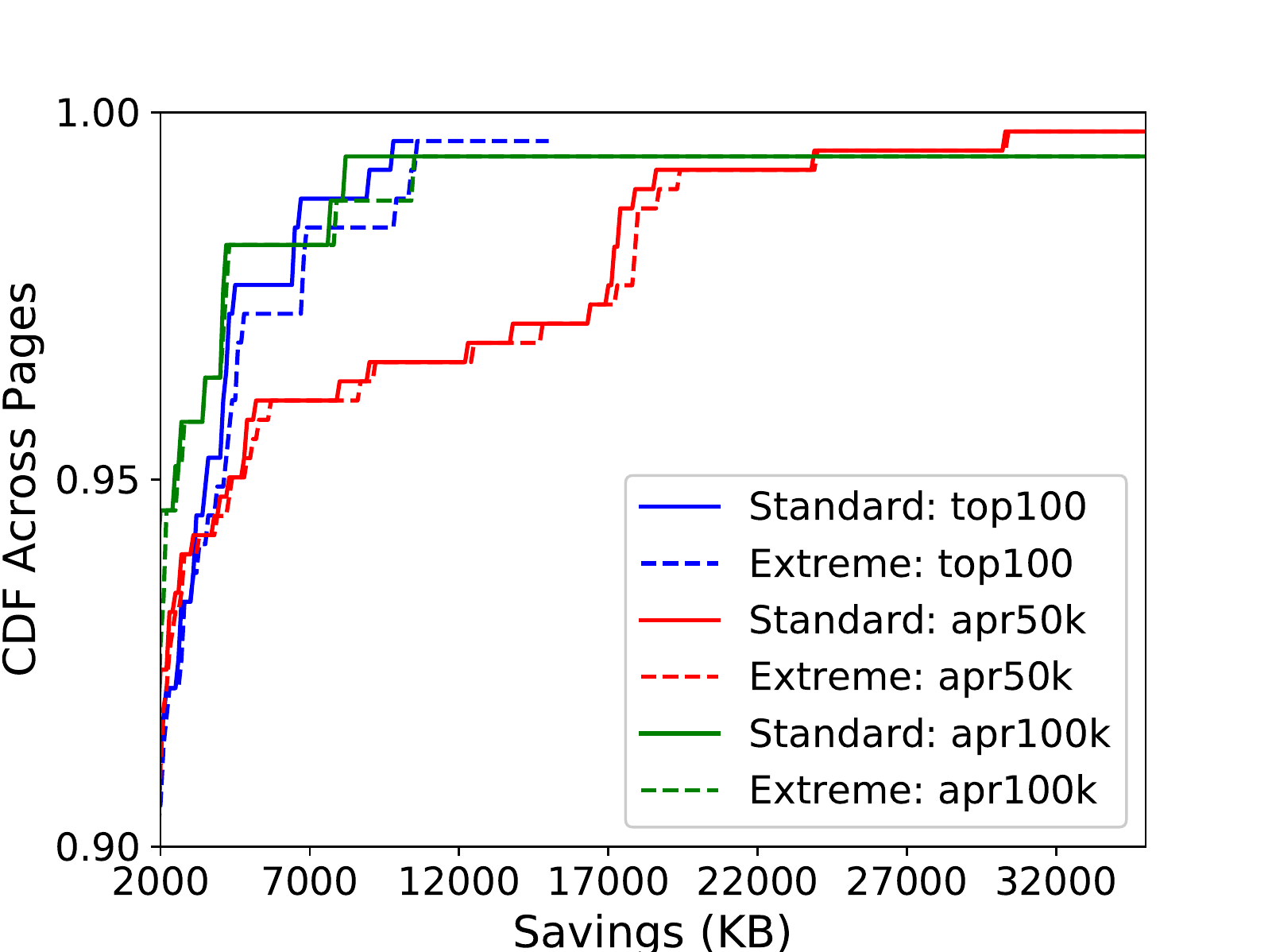, width=2.21in} \label{fig:raw-savings-and-ssim-b}}
   \subfigure[]{\psfig{figure=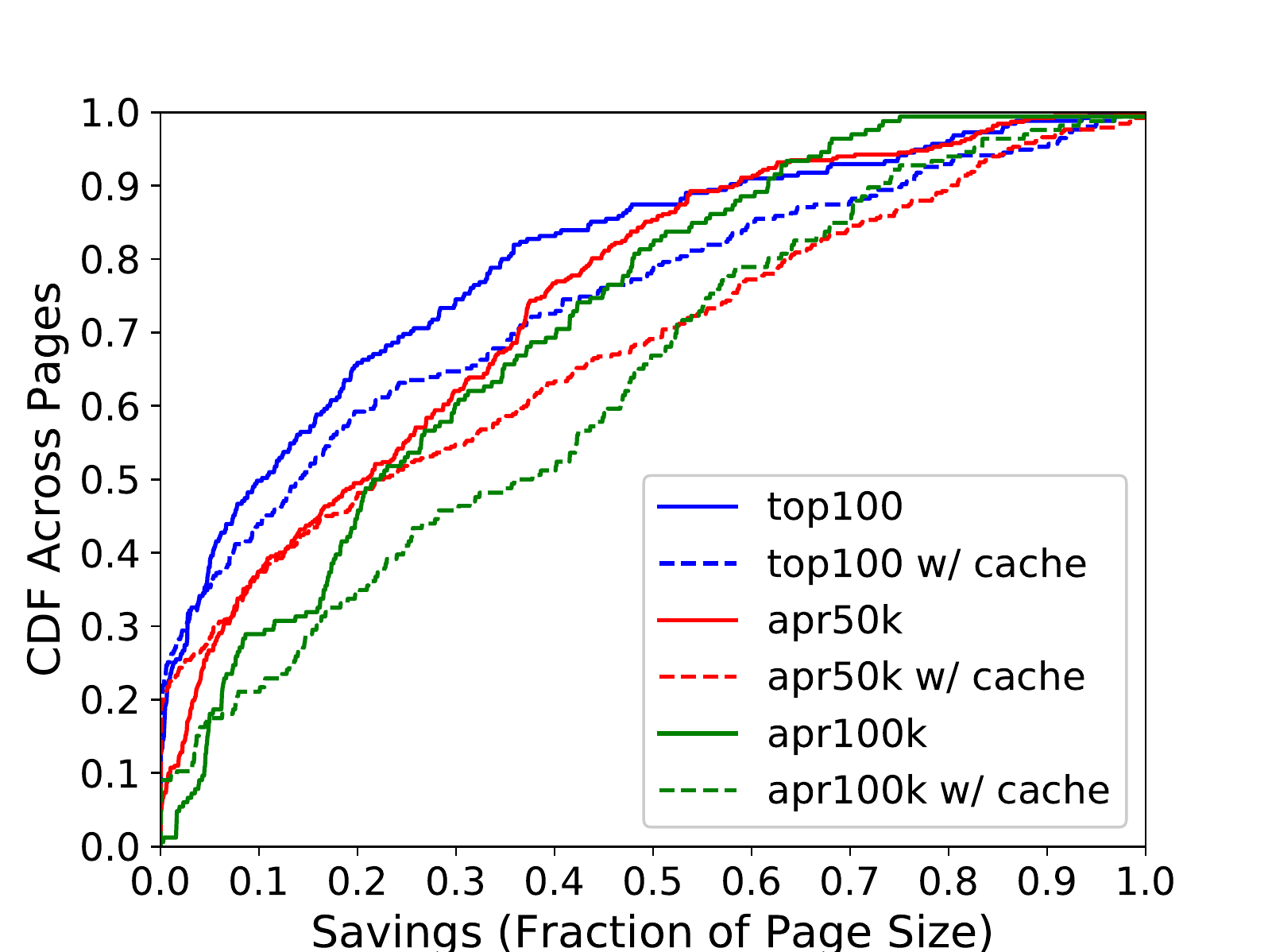, width=2.2in} \label{fig:raw-savings-and-ssim-c}}
   \caption{Room for bandwidth savings by adjusting image sizes. Shown in (a) and (c) are CDFs of potential raw and normalized savings respectively across pages of different ranks from our crawls. Normalization in (c) are against page weights with and without browser caching. As shown in (b), pages in the 90th percentile see room for over 7MB of savings.}
   \label{fig:raw-savings}
\end{figure*}

The research community has dedicated significant effort to design \textit{page load optimizations}. These works~\cite{Netravali-polaris, vroom-sigcomm-17, wprof-nsdi-2013, kelton-webgaze-2017} aim at speeding up the load times of Web pages by optimizing the order of Web object retrieval over the network. Such reordering schemes require server-side knowledge or assistance, and/or are unlikely to help in terms of bandwidth savings~\cite{Netravali2019-prophecy}.

Aside from load performance, data-savings is also a largely explored topic, with several commercial solutions already available (\eg ~\cite{opera-turbo, GoogleWebLight}). State of the art data-saving methods can be categorized as: 1)~middlebox transformations,~2)~server-side resource optimizations, and~3)~entirely client-side data saving approaches. In the following, we discuss each category in detail. 

\point{Middlebox Transformations} These methods rely on transparent HTTP proxies or middleboxes~\cite{Agababov2015-flywheel, Singh2015-flexiweb, li-nsdi-2016} to offer data savings via resource transformations. By operating transparently on path, they do not require server-side support. However, they cannot be used in presence of encrypted content (HTTPS) which is nowadays used by most websites~\cite{felt-https-2016,http-not-secure}. Popular resource transformations adopted by middleboxes include plain text \texttt{gzip} compression, image downsizing from its native dimensions to its rendered dimensions in the client's viewpoint, and content transcoding to formats which offer higher compression, \eg WebP. 

Flywheel~\cite{Agababov2015-flywheel}, Google's Web compression proxy, is perhaps the seminal work in the data-saving space. This proxy claims up to 80\% data savings via transcoding of image formats, and \texttt{gzip} compression of JavaScript and CSS.  FlexiWeb~\cite{Singh2015-flexiweb} is a follow-up implementation of Flywheel which leverages machine learning to optimize the trade-off between data savings and user quality-of-experience (QoE). Work from Alibaba~\cite{li-nsdi-2016} extends Flywheel's techniques to any mobile application by intercepting (unencrypted) mobile  traffic. 

Given the near-ubiquitous adoption of end-to-end encryption~\cite{felt-https-2016,http-not-secure} the potential for implementing such transformations via transparent middleboxes remains in question. Further, privacy is of concern to these methods, given that they require third party access to contents of webpages, much of which may be personalized. 

\point{Server-Side Resource Optimizations} 
These methods rely on some server side support to allow the clients to only fetch a minimalist version of the page~\cite{ChromeLiteMode, Netravali2019-prophecy, opera-turbo, azad-less-is-more, Chaqfeh-2020-cleaner, Goel2020}. The methods proposed include removal and reorganization of various Web objects from the page~\cite{ChromeLiteMode, Netravali2019-prophecy, opera-turbo}, detection and elision of unnecessary code~\cite{azad-less-is-more, Chaqfeh-2020-cleaner, Goel2020}, and URL rewriting based on content similarity to enable smarter and more effective caching at the client~\cite{Netravali2018-rc2}. To best understand which portions of pages to remove, or which content to rewrite, requires knowledge of the page state gained by completely loading the webpage before it is sent to the client, which is why these solutions require server-side support. 

To allow for data-savings without explicit server-side control, many of the above methods~\cite{Netravali2019-prophecy, opera-turbo} can be implemented as man-in-the-middle proxies which either break TLS or leverage URL redirection, \ie serve other Web page contents directly from their servers. For example, Google's Web Light~\cite{ChromeLiteMode} redirects the initial request for a webpage through its servers, without sharing the page's cookies to Google's servers. While this is a plus for privacy, as no client-side state can be inferred, it limits the reach of the approach, given that personalized content cannot be optimized~\cite{Tahir2020-deconstructing}. Further, such implementations leak information about which URLs are requested to third party servers, providing an opportunity to build full browsing profiles of end-users~\cite{Tahir2020-deconstructing}. Recent work~\cite{Kondracki-2020-meddling} has highlighted the above privacy risks of such approaches, and has also shown that many current implementations are built on outdated software, use substandard TLS certificate validation, and/or use weak TLS cipher suites, opening up users to additional security risks. 

Last but not least, these methods often make use of complicated rules for replacing/removing JavaScript~\cite{ChromeLiteMode, wang-shandian-2016, Goel2020, Chaqfeh-2020-cleaner}. Other solutions label dead code based on offline randomized user input testing~\cite{azad-less-is-more}. This implies that the efficacy of such solutions in terms of \webcompat remains quite uncertain and hard to measure, often requiring much manual effort to quantify~\cite{mozilla-webcompat}. Furthermore, when pages do break, there is often backlash from users and Web developers alike due to the lack of transparency of these systems~\cite{google-lite-break-pages}. 

\point{Client-Side Only Solutions} These methods run fully in the client, with no server support (either directly or via redirection) or support of on path middleboxes. While this approach offers the highest privacy guarantees, it is limited in which data-saving strategies can be adopted, since the actual contents of Web pages are unknown to clients until retrieved.

\textit{Content blocking} is the most common client-side data saving solution. This strategy simply blocks resources which can be identified as non-useful to users at the time of their request, such as advertisements in the case of ad-blocking. While more complicated solutions that block potentially useful page components, \eg JavaScript, are available~\cite{Chaqfeh-2020-cleaner}, they require apriori knowledge of page contents gained from observing the page load over a period of time.

Content blocking is also made available by Chromium under its Data-Saving mode to block all Web page images~\cite{blink-image-replacement}, replacing them with a single placeholder image. Image blocking saves users data, but it has drastic impact on the user experience of Web pages. While users are allowed to download these images individually, they have little to no context as to which images may be important to them. \ourwork balances data-savings and the user's browsing QoE; further, no user action is required. Our user study using \ourwork revealed that users tend to rate pages without images as completely broken (1 on a 1-5 scale at the median), whereas pages with data-optimized images received much more favorable responses (3 at the median). 

The work in~\cite{www-12-how-far} outlines what can be done from the client in terms of page load optimization, but focuses mainly on speculative caching (similar to ~\cite{Netravali2018-rc2}), as well as content pre-fetching to improve \textit{latency} rather than data-savings. While caching clearly reduces data sent over the network, it does not offer data savings under cold connections, which the same work stresses the prevalence and importance of~\cite{www-12-how-far}. \ourwork\ is primarily designed to save users data under cold caches, though our results (Section~\ref{sec:results}) highlight how \ourwork will not waste data in presence of hot caches. Further, recent works have shown security implications for too liberal caching policies, such as enforcing the caching of similar content between domains~\cite{sy-tls-2018}. 
\section{Potential for Image Savings}
\label{sec:measurement}

We begin our measurement by determining the potential for savings made available through less private, middlebox based optimization of images. Our analysis serves the purpose of creating a baseline for which to compare the data-savings of our more private, exclusively client-side, \ourwork. While the weight of images across webpages is known to be quite high (44\% as per HTTP Archive)~\cite{http-archive-almanac}, it is unclear how much of these images could be \textit{saved} using the middlebox approaches of image resizing, quality compression, and transcoding (see Section~\ref{sec:related}).

\point{Methodology} We resort to \textit{Web crawling}~\cite{urban2020beyond} to collect a representative dataset on the current status of image usage in the wild. To obtain a set of domains to crawl, we use the Majestic list~\cite{majestic_million} which contains the top million domains with the most referring subnets. We chose the Majestic list, as opposed to the more popular Alexa list~\cite{alexa}, as it is a free alternative that is still exclusively based on Web browser traffic~\cite{imc2018toplists}. 

We crawl 3 buckets of webpage rankings from the Majestic list (\topHun, \aprFty, and \aprHun)~\cite{majestic_million}. For each bucket, we select the first 100 websites, \eg pages 100-200 for the \topHun and pages ranked 50,000-50,100 in the \aprFty bucket. Given the importance of covering internal pages~\cite{ageel-imc-2020, urban2020beyond}, we crawl the first 10 links to the same domain, if available, from each top level domain. Our crawls originated a data-set encompassing $\sim$1.2k (300 landing and 880 internal pages) with, on average, 3 internal pages per domain. 

To perform the crawls by scripting Chromium version 83 using a combination of tooling via Lighthouse~\cite{ligthouse} and Puppeteer~\cite{puppeteer}. We use Lighthouse to load a page and collect various network and in-page statistics, such as network bytes associated with each resource requested, and the final rendered locations and sizes associated with all images from the Cascading Style Sheets (CSS) of the webpage. We use Puppeteer to obtain the response bodies of all network requests, and to scroll through the page to capture information of images which may be lazy loaded, or those added only in presence of user interactions. We ensure each webpage (landing and internal) is loaded using a cold cache. However, we also save the necessary HTTP headers (i.e. \texttt{Cache-Control, Expires, Last-Modified, and Etag}), to implement browser logic for determining whether a given request is cacheable~\cite{http-archive-caching}. This allows us to simulate the load of these pages with caching enabled, offline after our crawls. This analysis is particularly important for internal pages, where many resources may be cached from the landing page. 

Lighthouse currently provides rough estimates of wasted bytes due to a page failing to implement standard image data-saving optimizations (See Section~\ref{sec:related}. For our measurements, we extend these estimates in two ways. In the following, we detail both extensions to Lighthouse reports and, for each, provide an analysis of the extensions based on the $\sim$1.2k websites from our crawl. 

\begin{figure}[tb]
    \centering
    \includegraphics[width=2.5in]{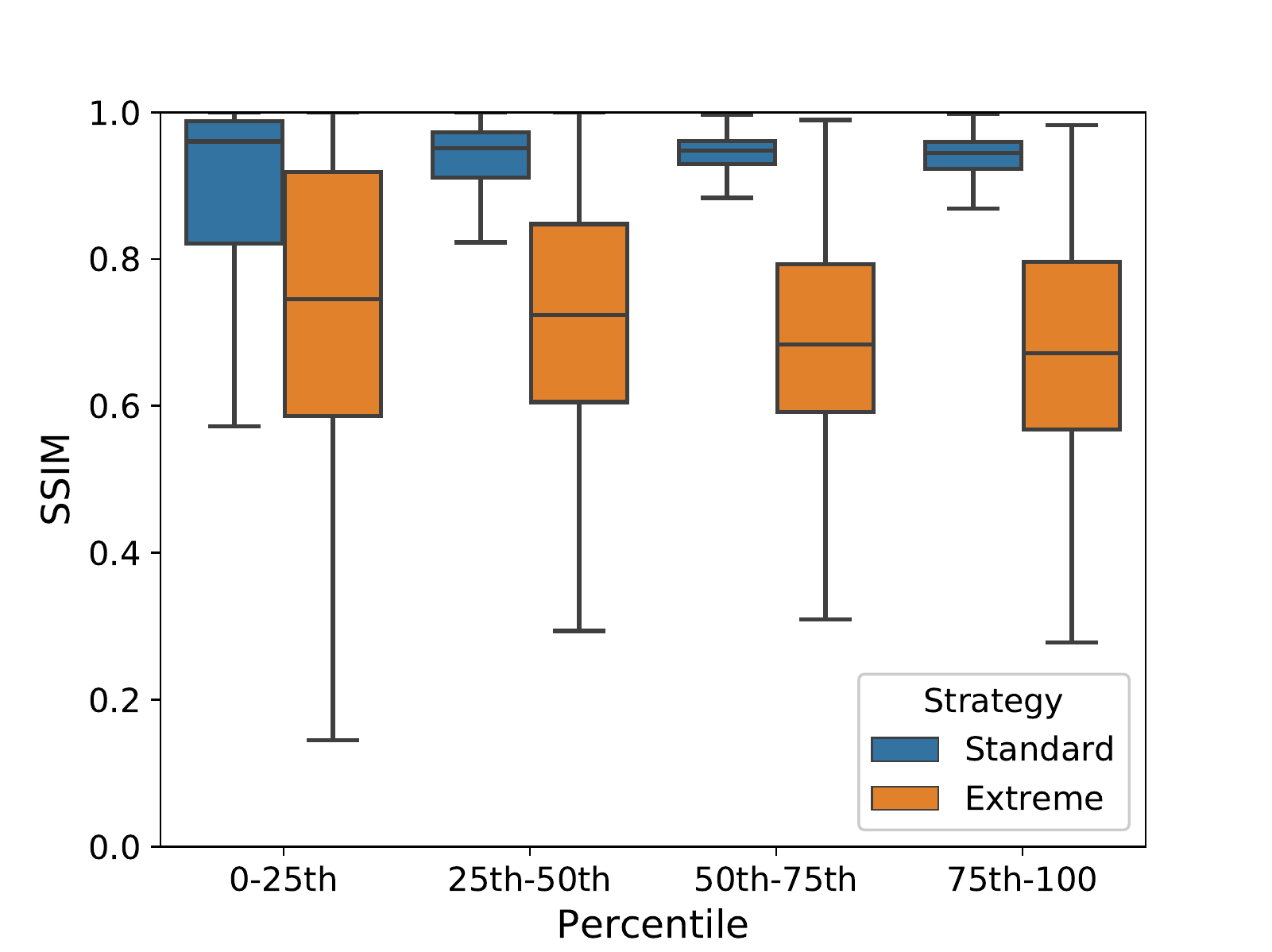}
    \caption{The resulting trade-offs in image quality, measured via SSIM, across the different levels of potential savings from Figure~\ref{fig:raw-savings}.
    }
    \label{fig:savings-ssim}
    \vspace{-.2in}
\end{figure}

\point{Image compression pipeline} We estimate the potential data waste in images by manipulating all the HTTP response bodies of image requests through 3 image optimization techniques as recommended by Lighthouse and employed by proxy based approaches (\eg Flywheel~\cite{Agababov2015-flywheel}), that is, image resizing, quality compression, and transcoding. Our first extension to Lighthouse data-saving measurements is to pipeline images through these 3 optimizations, as Lighthouse currently only applies these individually. This approach can underestimate savings as these optimizations compound to save data~\cite{Agababov2015-flywheel, Singh2015-flexiweb}.  
We employ two versions of the pipeline, \textit{standard} and \textit{extreme}, which have trade-offs in terms of savings versus potential impact on the user's QoE. To quantify the reduction in QoE caused by image savings, we use the \textit{structural similarity metric}~\cite{ssim-metric} (SSIM), a full image reference metric which is commonly used to measure quality degradation in images due to transformations (blurring, compression, color reduction, etc.). 

For the resizing component of the pipeline, in standard mode, we resize the image as sent over the network to its CSS attributes width and height; for extreme mode we resize the image to half of these values. We note that the CSS attributes depend on the size of the viewport, and hence smaller viewports may achieve higher relative savings for the same image at the same perceptual quality. In all our experiments, we instrument Chromium to emulate a Pixel 2 which has viewport size of 411x731 (5.5 inches), or the most popular size in 2019~\cite{mobile-phone-stats}.
Following image resizing, all jpeg, tiff, png, bmp, and gif images are transcoded to WebP, a ``next generation'' image format which offers higher compression with visual quality comparable to the other formats~\cite{google-webp-study}. In standard mode WebP images are compressed by reducing their quality setting to 85 (out of 100), as this is reported as the best trade-off between savings and SSIM degradation according to previous works~\cite{Agababov2015-flywheel, Singh2015-flexiweb}. In extreme mode, we aggressively reduce image qualities to 10 (out of 100). 

Results from the image compression pipeline are shown in Figure~\ref{fig:raw-savings} (a) and (b). Pictured first is the CDF of savings, in KBytes, for images across pages from the \topHun and \aprFty websites. 
What we can observe is that lower ranked pages see generally higher savings, with medians of 152KB, 292KB, and 308KB for standard mode of \topHun, \aprFty and \aprHun respectively, and 189KB, 375KB, 390KB for extreme mode on the same. This result is intuitive, as more popular pages are expected to be more optimized. 
While the savings offered for the median page is rather modest, the distribution of savings is quite long-tailed, as shown in Figure~\ref{fig:raw-savings} (b). The figure shows that even top ranked pages see savings of over 3MB at the 95th percentile. Further, the gap between the standard and extreme modes is modest, being at most 100KB at the median page across the ranks. This is likely because savings offered through resizing and reducing quality begin to diminish when configured further past that of our standard level. 

Outside of raw bytes, we compare savings as a fraction of the total page weight. Figure~\ref{fig:raw-savings} (c) provides the normalized savings, for our standard mode, across our crawls broken up by page rank. As before, the lower ranked pages offer more potential savings, with 10.3\%, 20.1\%, 21.9\% of the median page being saved for \topHun, \aprFty, and \aprHun respectively. This jumps to 30.8\%, 38.9\%, and 44.5\% for in the 75th percentile across the respective crawls.

We also analyze the savings normalized by the weight of pages under caching. Specifically, we do not count URLs (including images) that are a) found on both top level pages and second-level pages, and b) marked as cacheable by their HTTP headers~\cite{http-archive-caching} towards the total weight and savings of second-level pages. Note that we assume a double-keyed HTTP cache~\cite{implement-dual-key, davis-safari-caching}. It follows that when determining repeat URLs, the same image on different domains will be counted for savings across both pages. Figure~\ref{fig:raw-savings} (c) also shows this result, where we can observe that savings increase in the presence of warm caches; specifically savings of up to 14.2\%, 21.6\%, and 36.8\% are observed for the median page in \topHun, \aprFty, and \aprHun respectively. This is because many Web resources, other than images, are shared between landing pages and inner pages, thus increasing the relative weight of data-saving techniques. 

Moving to SSIM analysis, we observe that standard mode provides rather mild QoE trade-offs, while the trade-off for the extreme mode is harsher. Figure~\ref{fig:savings-ssim} shows paired boxplots of the resulting SSIM metrics of images, optimized by both standard and extreme modes and bucketed by their resulting percentile of data-savings. While the median SSIM of images for all levels of savings in standard mode does not fall below .95 (.83 for the 25th percentile), the median of extreme mode sits at .74 across all levels of savings. Finally, we also observed that images that benefit more from data-savings see lesser reductions to their visual quality. This analysis suggests that high data-savings are possible with modest QoE impact, with standard mode offering the more preferable trade-off.

\point{CSS sprites} Lighthouse ignores potential savings from resizing images which are embedded by CSS, or CSS background images. This is because a typical use for background images is CSS sprites~\cite{css-sprites}, or images that consist of multiple smaller images embedded in one parent file (or sprite sheet). The sprite sheet is then dynamically cropped by the webpage to render the component images. Measuring savings for CSS sprites correctly requires accounting for the final size and location of all sprites, not just the original image. Due to this complexity, Lighthouse currently ignores calculating data-savings for \textit{all} background images, thus leading to potential savings underestimations.

\begin{figure}[tb]
   \subfigure[]{\psfig{figure=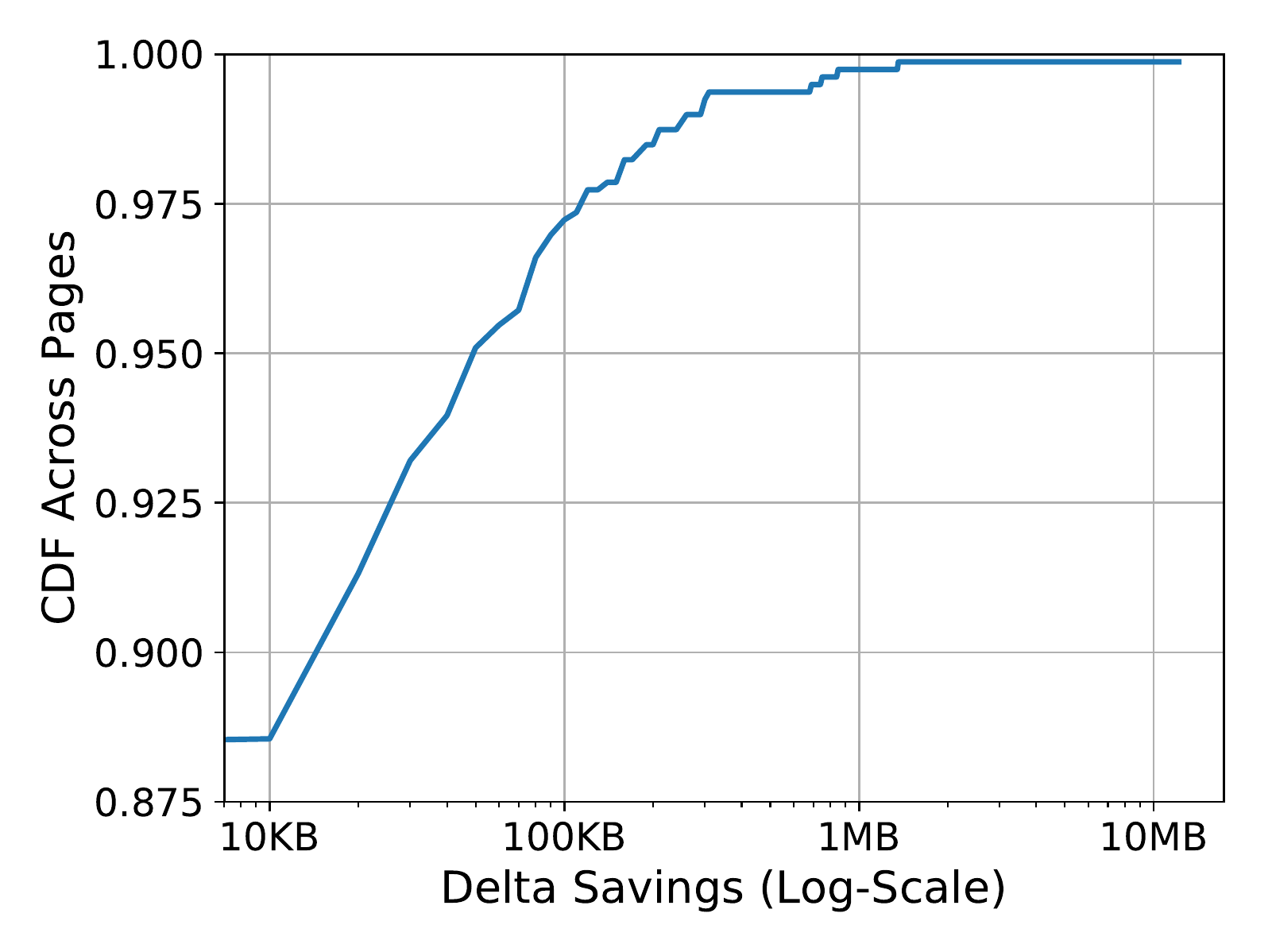, width=1.63in}\label{fig:sprite-diff-a}}
   \subfigure[]{\psfig{figure=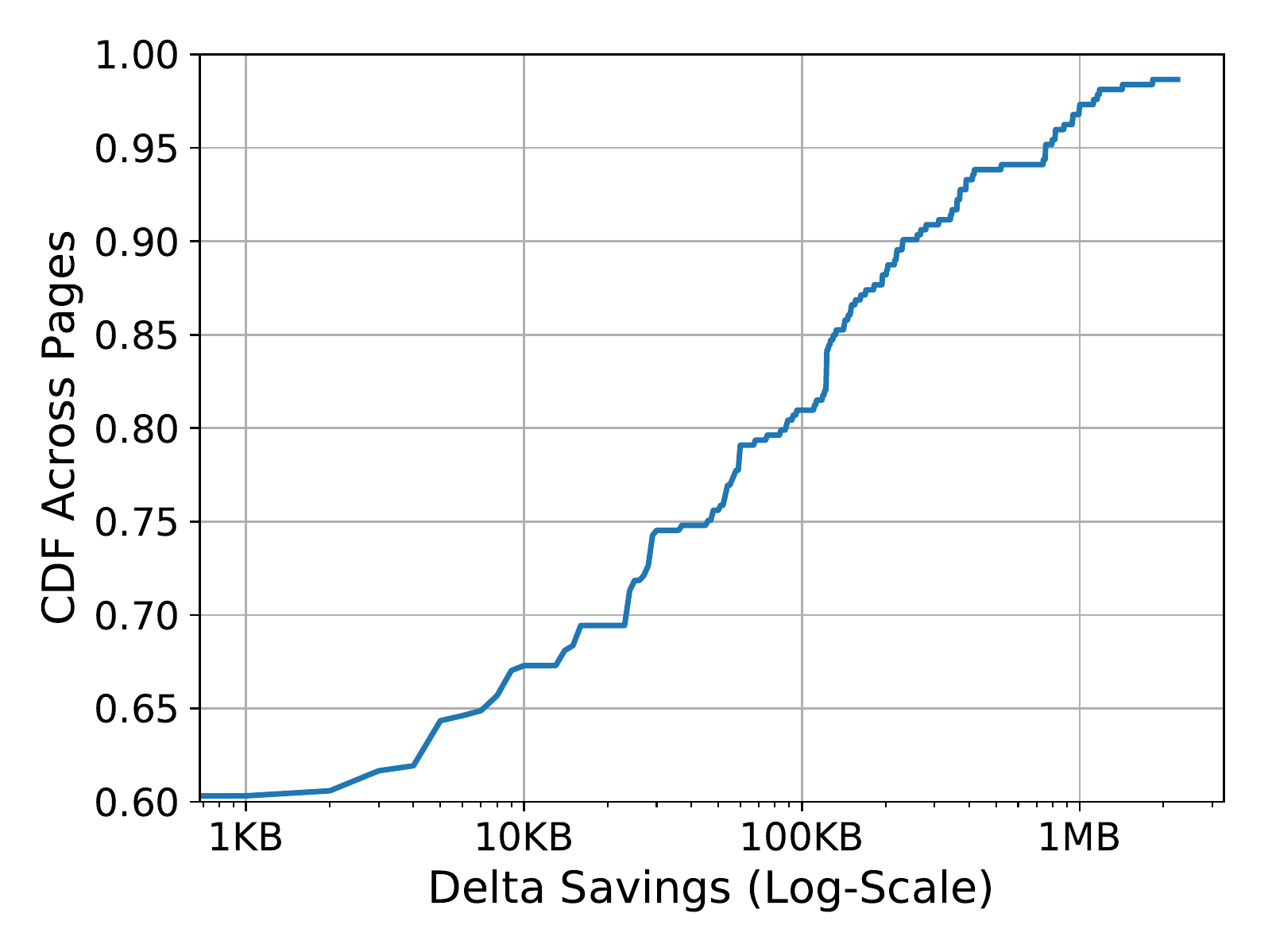, width=1.63in} \label{fig:sprite-diff-b}}
   \caption{Log scaled difference in savings (KBytes) when (a) not considering CSS background images and (b) when failing to appropriately calculate savings for CSS sprites.}
   \label{fig:sprite-diff}
 \vspace{-0.5cm}
\end{figure}

We identify CSS sprites, or more generally any images that are cropped dynamically using the CSS property \textit{background-position}, and separate these from normal CSS background images. We compute savings for normal background images using the pipeline previously discussed in this section. To calculate savings for CSS sprites, we compare the total area used by sprite sheets on the page with the total area of these images as sent over the network. Figure~\ref{fig:sprite-diff} (a) shows the Cumulative Distribution Function (CDF) of the data-savings which are currently missed by Lighthouse due to ignoring background images. Figure~\ref{fig:sprite-diff} (b) shows a CDF of the overestimation in savings seen from incorrectly resizing entire CSS sprite sheets to their component sprite size. The CDFs refer to the full set of pages from our dataset, and show the change in KBytes of savings in log scale. The graphs start at (a) 0.875 and (b) 0.6 for visibility as only~$~$12.5\% and $~$40\% of pages were affected for (a) and (b) respectively. While not accounting for background CSS images only misses $<50~KB$ of savings for 88\% of pages, there are~5\% of pages for which $>100~KB$ are missed. Resizing the CSS sprite sheet to the size of a single sprite can cause an overestimation of~$100~KB$ and $2~MB$ of savings for the pages at the~60th and~90th percentile, respectively.

\section{Implementation}
\label{sec:impl}
Measurements from our previous section are motivating evidence that image optimizations contribute to significant bandwidth savings for Web browsing, especially in the upper tail of lesser-optimized pages which see savings on the order of MBytes. However, such savings represent a bound for image savings attained by either privacy invasive approaches, \eg using middleboxes, TLS interception, URL rewriting, or solutions which require some form of server collaboration. 

In this section, we introduce \ourwork, a collection of techniques which realize image data-savings directly at the client (\eg a browser), thus offering higher privacy guarantees. \ourwork consists of two main techniques, \rewriting and \fetching, which are both described in the following.  

\subsection{Image Server Instrumentation from the Client}
The key challenge for a client-side data saving approach is that it cannot apply transformations in the same way as middleboxes; when images are received by the browser, the user's data has already been wasted. Instead, it requires a way to ask the server for a more compressed version of images. Our intuition is that this is possible thanks to the recent widespread of \textit{image services} run by popular content delivery networks (CDNs) like Fastly~\cite{fastly}, Akamai~\cite{akamai}, and  CloudFlare~\cite{cloudflare}. These image services offer similar means as middleboxes to reduce image file sizes. Such savings are typically made accessible by configuring parameters in the URL of the image being served by the CDN. 

However, these image services are not always configured optimally. For one, images uploaded to these services may be resized in a ``one-size fits all'' manner, for convenience, even though mobile pages, for instance, are accessed from a large diversity of device types and screen sizes. For smaller devices, this implies images can be further resized without observable quality degradation. Further, browser fragmentation means modern image formats (\eg WebP and AVIF) are not always supported. For these reasons, image services can be configured in an overly conservative manner to not deliver these modern formats, missing out on significant savings (see Figure~\ref{fig:raw-savings}). Lastly, image services may be configured to deliver images under higher visual quality (as determined, for example, by SSIM discussed in Section~\ref{sec:measurement}), rather than higher data-savings. However, many image formats are able to keep much of the same visual quality at even a significant level of compression, \eg jpeg and WebP images provide no noticeable visual quality reduction at 85\% compression levels~\cite{Agababov2015-flywheel} (See Section~\ref{sec:measurement}). As we will show later (see Table~\ref{fig:rewrite-results}), these configurations of image services can miss up to 70\% savings across real pages.

We design \ourwork\ to detect the use of such image services, and to uncover potential data-savings in their configurations. Specifically, \ourwork\ detects whether or not an image server supports the same transformations as the standard mode outlined in Section~\ref{sec:measurement}, that is, CSS right-sizing, quality reduction, and format transcoding. If such support is detected, \ourwork modifies HTTP requests for these images in real-time to automatically apply such transformations, thereby optimizing bandwidth consumption. We call this component of our work \rewriting.

To identify the presence or lack of an image service associated with a given image, \ourwork searches for parameters in the URL of images that might related to the image's dimension, compression level, and format. For example, in the URL\footnote{\texttt{\url{https://static.wixstatic.com/media/ 98a2de_37749ccfe79f48d1a977af77d1c2bd0e~mv2.jpg/v1/fill/ w_400,h_52,al_c,q_100,usm_0.66_1.00_0.01/Classic_Car_Painting_By_Pavel_Hol\%C3\%BD.jpg}}}, the parameters \texttt{w\_400}, \texttt{q\_100}, and extension \texttt{.jpg} correspond to the actual dimensions, quality compression level (out of 100), and format of the downloaded image. The equality of the image data and the URL parameters suggests that the image can be dynamically resized, compressed, and transcoded just by changing such parameters on the fly. 

Editing URL parameters is not without risk, as a URL may simply be statically defined with no image service available, and should thus not be edited. This manner of false positive can, at the very least, cause an extra unnecessary round trip and mitigate bandwidth savings, and at the worst actually hurt bandwidth savings.

To avoid latency and bandwidth harming requests as much as possible, \ourwork takes a two step approach to \rewriting. First, we rewrite any value in the URL that matches a native size, quality, or format property of the image. Intuitively, this method achieves high true positive rate, but also high false positive rate. Thus, second, we generate a series of rules from the true positives to increase precision; instead of rewriting any location in the URL matching a property, we only rewrite URL parameters matching such rules (\eg \texttt{w\_} in our example above).  We further extend these rules by manually exploring mobile vs. non-mobile versions of the page and the image service APIs of 12 popular CDNs\footnote{Rules and manual efforts found in the document: https://tinyurl.com/86srq141.}. We create such rules across the $\approx$10k images obtained from our crawls outlined in Section~\ref{sec:measurement}. In the end we chose a subset of rules for matching which gave the best trade off in terms of true positive and true negative rates for all our observed images.

Figure~\ref{fig:rewrite-results} summarizes the final results of the \rewriting process. We can observe that a total of~50 unique rules were utilized which affect~16.1\% of images for an error rate of~7.2\%. Breaking down the error rate,~6.6\% of images returned a HTTP~404 status code, implying a need for a single re-fetch when running \ourwork. The latter~0.6\% returned a size greater than the original size. Overall, of the affected images,~69.9\% of the original image size was saved on average. We discuss the impact \rewriting has on page data-savings in the following section.

\begin{figure*}[t]
\centering
   \scriptsize
   \centering
    \resizebox{1\textwidth}{!}{\begin{tabular}{l r r r r r r}
     \toprule
      \bf Image Optimization & \bf Unique Rewrite Rules & \bf TPR & \bf FPR (404) & \bf FPR (No Savings) & \bf Savings\\
     \midrule
     Right-Sizing & 39 & 15.4\% & 6.1\% & 0.6\% &  66.2\% \\
     Quality Reduction & 5 & 3.7\%  & 1.2\% & 0.01\% & 53.7\% \\
     Format Transcoding & 6 & 3.9\%  & 0.1\% & 0.01\% & 44.1\% \\
     Any Transformation & 50 & 16.1\%  & 6.6\% & 0.6\% & 69.9\% \\
     \bottomrule
         \end{tabular} 
          
    }
     \caption{Quantifying the effectiveness of instrumenting image services for better image optimizations from the client. Savings is in terms of average reduction in size of these images. While a small fraction of pages are able to be rewritten in this way, the relative savings is quite large.}
    \label{fig:rewrite-results}
\end{figure*}

\subsection{Fetching Less with Range Requests}
\label{sec:impl:range}

The HTTP standard outlines the ability to request arbitrary bytes of an HTTP response over the network. This is achieved by an HTTP \textit{range} header being attached to a request which indicates the amount of bytes of the resource that should be sent from the server. Assuming this capability is properly implemented at the server, a \texttt{206} status is returned for the request, with the subset of bytes requested as the response body. As Web image formats were designed to display render even under slow or lossy networks, not only can images be partially requested, but major image codecs (\eg libpng, libjpeg, libwebp) support partial rendering as well. As~96\% of the servers in our crawls supported range requests, our rationale is to combine the use of range requests with \rewriting to achieve savings on a more general set of webpages. We call the process of \ourwork to request less for images its \fetching component. In Section~\ref{sec:results} we compare the effectiveness of the combined \rewriting and \fetching components of \ourwork to existing approaches, \eg middleboxes and Google Web Light~\cite{GoogleWebLight}, in terms of data-savings.

While clearly requesting 50\% of the bytes of an image can usher in 50\% data-savings, what remains is the impact on semi-complete images on user QoE. Generally, the first X\% of the bytes of Web images can be used to render the top most X\% of the pixels of images. This implies that only the top half (approximately, given compression) of the images will be displayed given a 50\% range request, making most images appear broken.

However, there are some factors which can alleviate this impact. First, \textit{progressiveness} is a rendering mechanism of jpeg and png formats in which image data is encoded such that it is able to be rendered in \textit{layers} as opposed to top-to-bottom. What this implies is that an image can be rendered in its entirety, albeit at a lower quality (SSIM), given only a fraction of the available bytes. As discussed in Section~\ref{sec:measurement}, much of the potential data-savings for images come from the fact that they are sent at a larger size than they will be rendered at the client. Due to downsizing, when they arrive at the client certain progressive images can be rendered at high quality even with a fraction of their payloads requested. 

For non-progressive images, when applying \fetching, \ourwork performs a visual trick to make pages appear less visually broken while still attaining meaningful savings. This technique takes the partial data of the image that is obtained over the network, and fills in the broken gaps of empty space with reflected and blurred content of the image, which we call \textit{reflections}. The idea stems from a popular technique on the Web known as \textit{image previews}~\cite{smashing-image-previews}. This is a technique employed by many popular Web services (\eg Facebook~\cite{facebook-previews} and Medium~\cite{medium-image-previews}) where pages display small and blurred portions (on the order of bytes) of images before the full versions are downloaded, as opposed to empty spaces or placeholders. However, since \ourwork does not have access to server-side control, and thus full image data, we cannot pre-process the images offline to make previews. Instead, we use the partial data in the range request to make reflections on the fly, at the client.

\begin{figure}[bth]
   \centering
   \subfigure[]{\psfig{figure=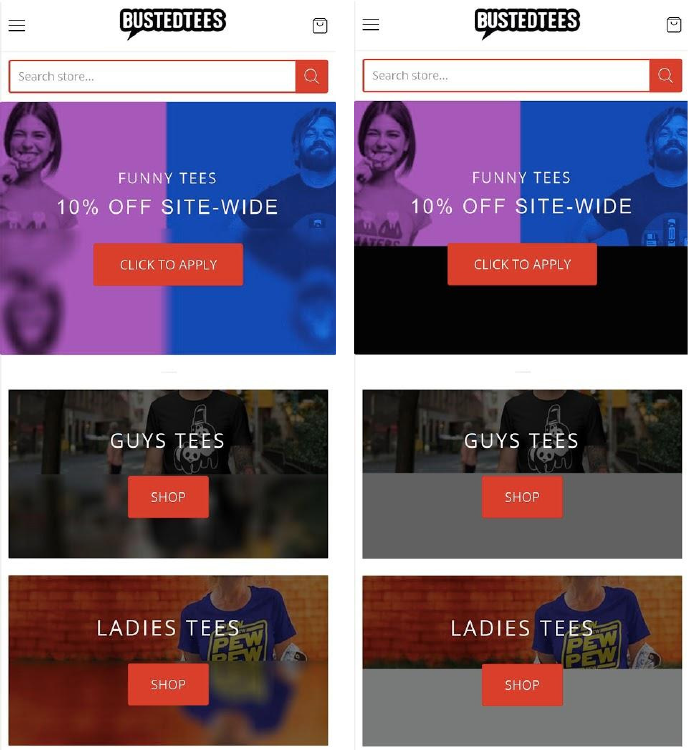, width=2.2in}
   \label{fig:demo-reflection-progressive-a}}
   \subfigure[]{\psfig{figure=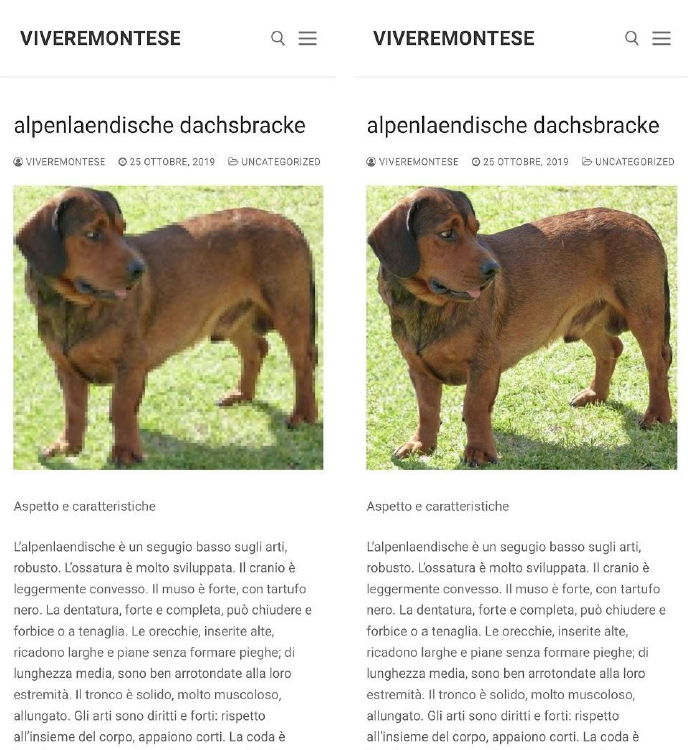, width=2.2in} 
   \label{fig:demo-reflection-progressive-b}}
    \caption{The visual completeness of (a) reflection and (b) progressiveness of images. The left pair shows the state of the page with 50\% of the image data requested. The page with reflection is still 93\% visually complete while the page without is only 78\% virtually complete. The right pair shows the state of a page with a large progressive image. The page is already 95\% visually complete with only 15\% of the data requested (99\% with 80\% requested).}
   \label{fig:demo-reflection-progressive}
\vspace{-7mm}
\end{figure}

Figure~\ref{fig:demo-reflection-progressive} shows a visualization of \fetching in case of regular (a) and progressive images (b). For Figure~\ref{fig:demo-reflection-progressive} (a), both images show the page with 50\% the image data requested. When reflection is applied (left most), the page is 95\% \textit{visually complete} (according to the SpeedIndex metric, see Section~\ref{sec:results:user} for more details) and only 73\% visually complete without reflection. Figure~\ref{fig:demo-reflection-progressive} (b) shows a page with 15\% (left most) and 80\% (right most) of the image data requested, respectively. Given the progressive image is sent over the network at dimensions much larger than its final rendered ones, the page is still 95\% visually complete with only 15\% of the data fetched. 

\vspace{-0.25cm}

\subsection{\ourwork}

We implement \rewriting and \fetching as a Puppeteer~\cite{puppeteer} application. While we test with Chromium version 83, our application can function out of the box on any Chromium based browser or one in which supports the Chrome Debugging Protocol (\eg Brave, Edge, and Opera~\cite{cdp-support}). Further, while \ourwork is prototyped as an external application, many of the components use internal browser APIs. We discuss the potential for \ourwork\ to be fully integrated with the browser in Section~\ref{sec:discussion}.  

To perform \rewriting, \ourwork intercepts all HTTP requests associated with images as defined by the Chromium network stack. Each request URL is associated with a DOM node of the Web page which can be used to extract the image's CSS width and height as needed for \rewriting. Next, the actual request URL is run through a regex to reconfigure the URL parameters to fetch a lighter version of the image, if possible, in the same manner as discussed in Section~\ref{sec:measurement} and Table~\ref{fig:rewrite-results}.

Moving to \fetching, since \ourwork\ does not assume server cooperation, it does not know the file size of an image apriori, which is needed to form a range request. For this reason, \ourwork assigns a range header to instead request the first 2KB of every image. Contained in the server's response is the image's full size in bytes, and the \textit{metadata} for the image. This metadata is immediately passed to the browser since it is used to facilitate the final layout of the page~\cite{webdev-cls}, which should not be delayed. Once the full image size is known, a second range request is immediately issued to obtain the lighter version of the image as a fraction of its known total size. 

Following the request procedure, the resultant image is built in memory using the concatenated data from the first and second range requests to prevent wasted bandwidth. To display the image from memory, the image is in-lined in the Web page using a \texttt{dataURI}~\cite{data-uris} on the associated DOM element which was previously obtained. The metadata from the initial 2KB of the image is used to determine, on the fly, its progressiveness. If the image is progressive, the dataURI simply consists of the data requested over the network. If the image is not progressive, the image data is decoded, reflected, blurred, and re-encoded to a final dataURI. 

Finally, fallback cases are necessary for both  \rewriting and \fetching components of \ourwork. If the initial 2KB request returns a 404, then  \rewriting is aborted and only \fetching is used. In case of a false positive, \eg the returned image is bigger than the original, \ourwork\ can only proceed to \fetching. For \fetching, range request support is determined on the fly if the server does not respond with the expected initial 2KB, or does not return the expected response headers to notify \ourwork of the image's total size. In these cases, the full image, after having been subjected to \rewriting, is downloaded in its entirety. As discussed before, \ourwork only sees about 7\% false positives when \rewriting, and only 4\% of servers do not support range requests needed for \fetching. 
\section{Results}
\label{sec:results}

We move to evaluate \ourwork in terms of its impact on user QoE, data-savings, and page load performance. We also compare the data-savings of \ourwork to the less private middlebox optimizations of Section~\ref{sec:measurement} and Google Web Light.

\begin{figure}[t]
    \centering
    \includegraphics[width=2.5in]{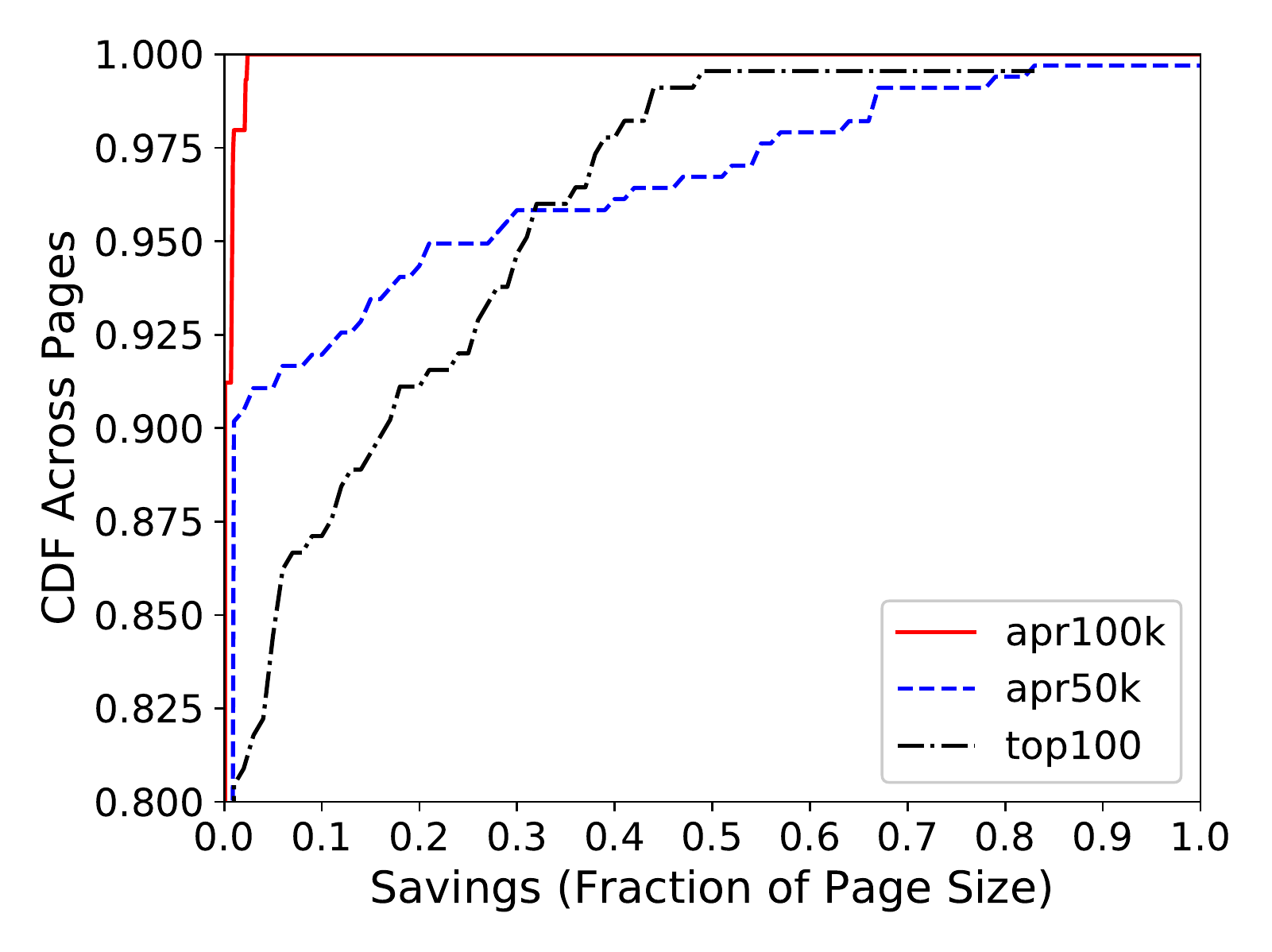}
    \caption{Page savings recovered by manipulation of server-side compression via \rewriting. As $\approx$16\% images are optimized, savings are shown from the 80th percentile.
    }
    \label{fig:url-rewrite-savings}
\end{figure}

\begin{figure*}[t]
   \centering
   \subfigure[]{\psfig{figure=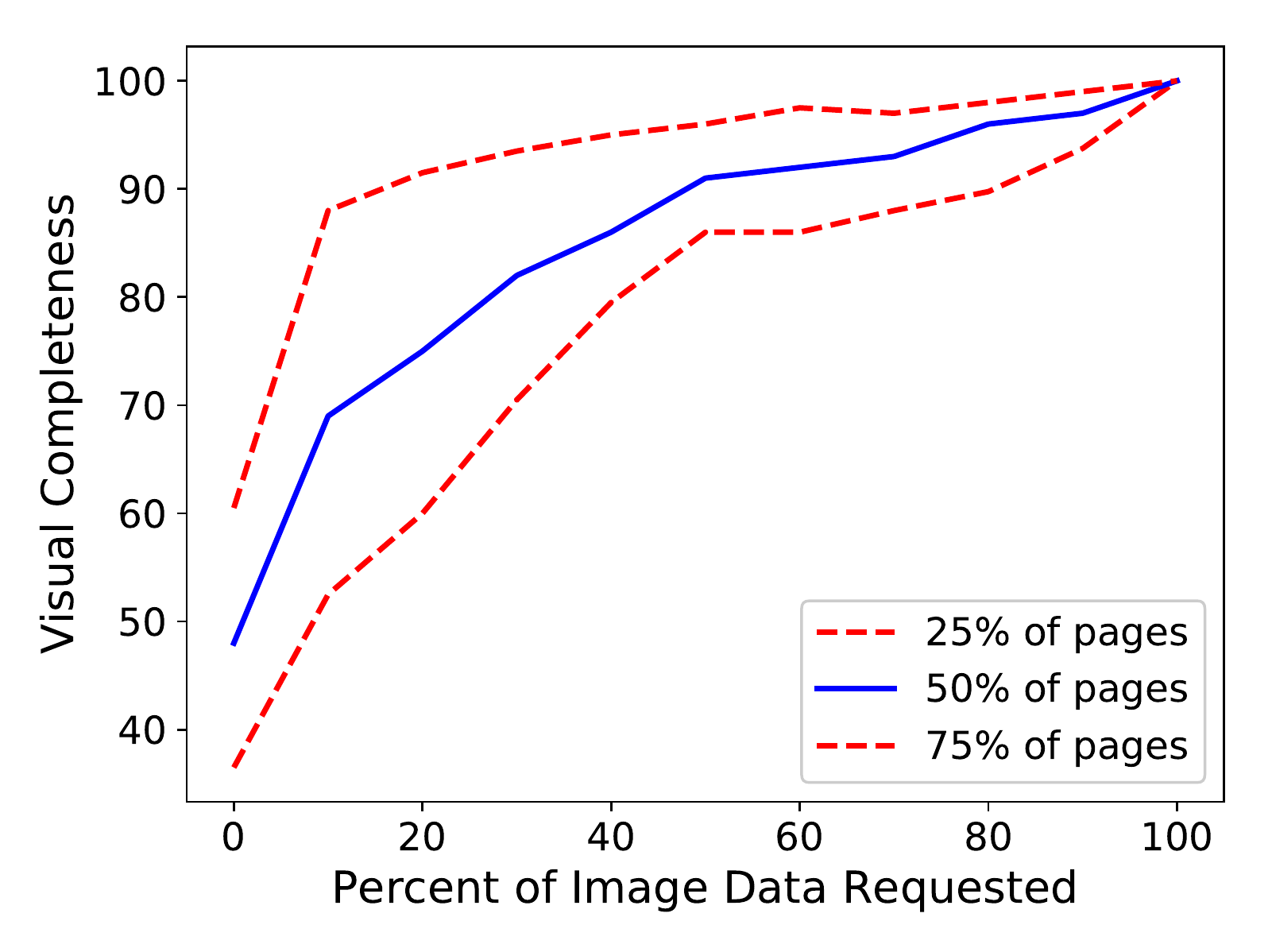, width=1.72in}\label{fig:range-request-results-a}}
   \subfigure[] {\psfig{figure=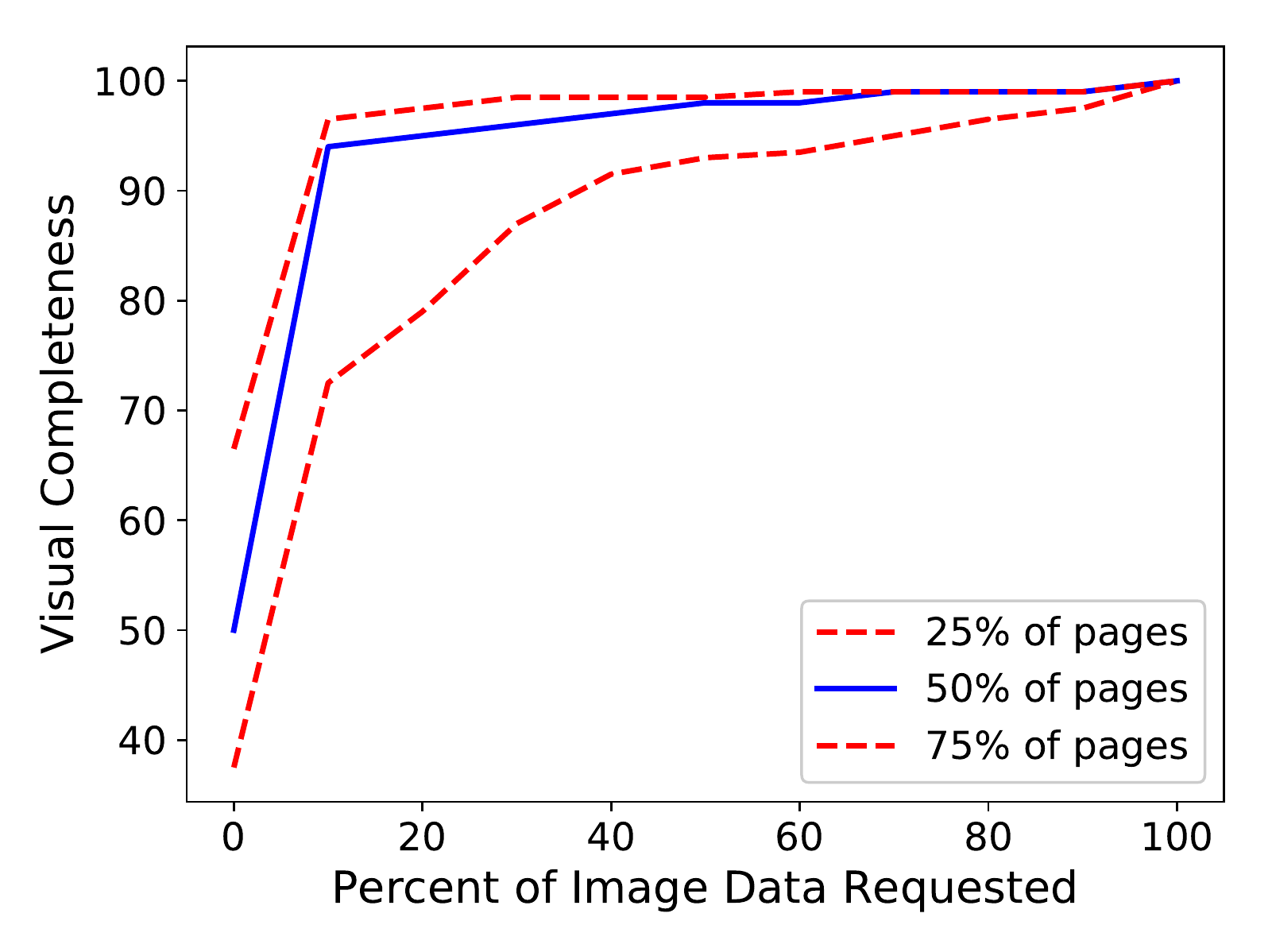, width=1.72in}\label{fig:range-request-results-b}}
   \subfigure[]{\psfig{figure=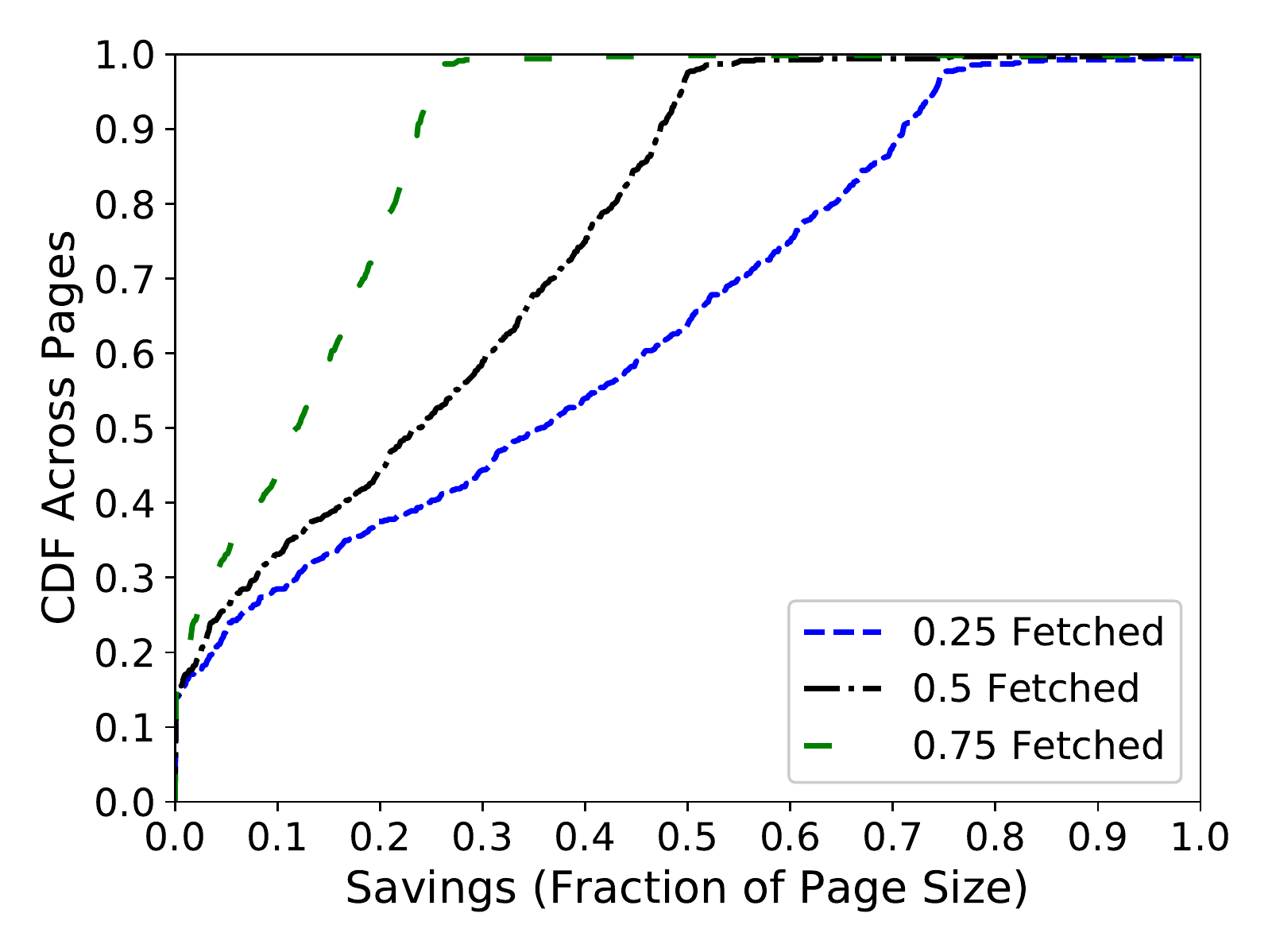, width=1.72in}\label{fig:range-request-results-c}}
   \subfigure[]{\psfig{figure=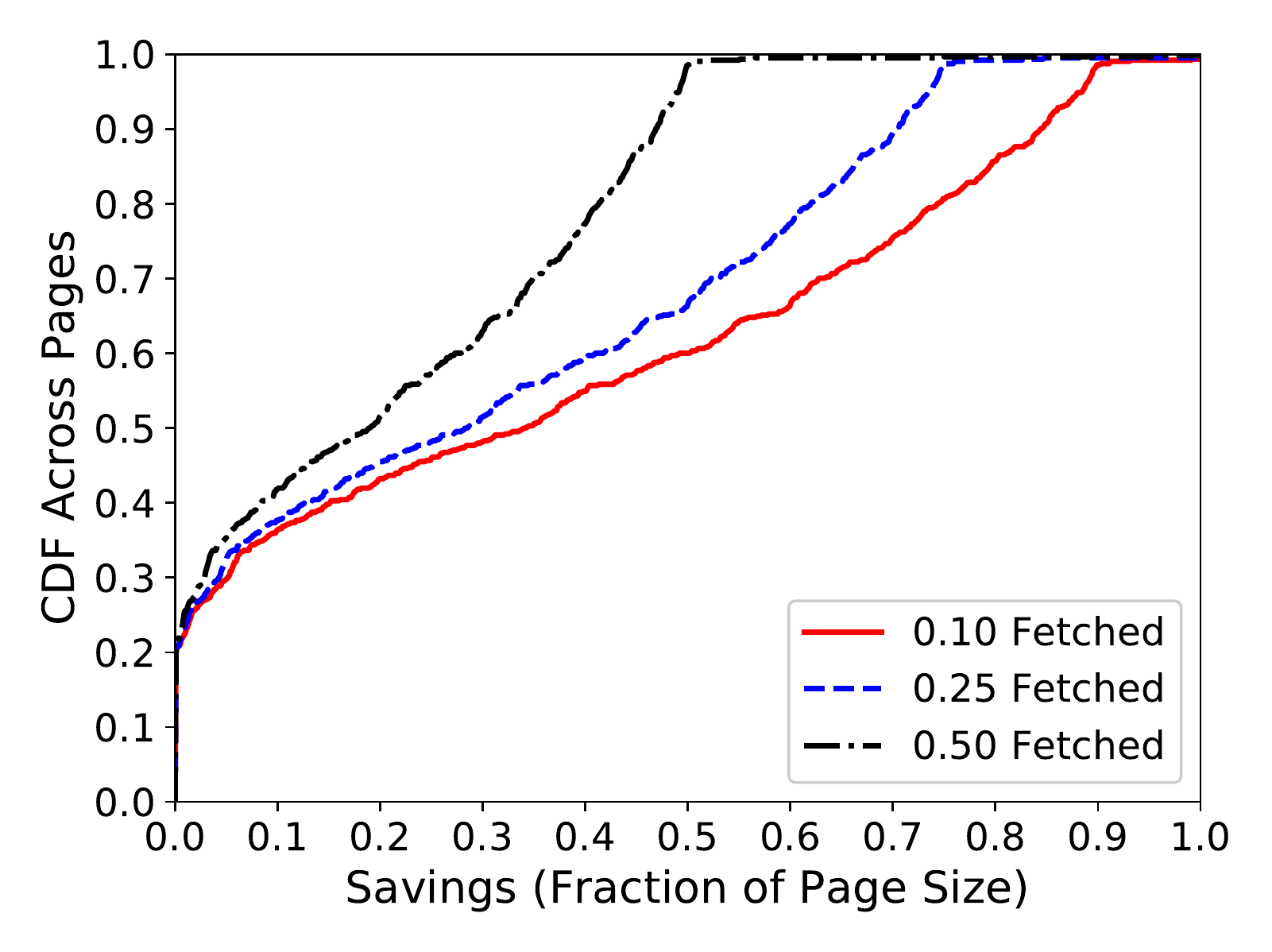, width=1.72in}\label{fig:range-request-results-d}}
   
   \caption{CDFs of savings offered by \ourwork as well as the tradeoff in visual rendering quality. Plots (a) and (b) quantify visual completeness values against the percent of image data requested. The plots picture all pages and only those with a supermajority of progressive images respectively. Plots (c) and (d) convey normalized savings for the same, broken up by three levels of image data requested. For (d) the levels show less data requested, since pages with progressive images are more visually complete with the same amount of data requested.}
   \label{fig:range-request-results}
\end{figure*}

\subsection{Bandwidth Savings and Visual Trade-offs}

We begin by analyzing the potential data-savings offered through \rewriting. Figure~\ref{fig:url-rewrite-savings} shows the CDF of the fraction of page bytes saved, per the URLs in each \textit{bucket} of our crawls (Section~\ref{sec:measurement}). The figure shows that \rewriting can only be used for~10--20\% of the URLs in the two higher ranked popular buckets (\topHun and \aprFty), while it has little effect on less popular URLs (\aprHun). This is because higher ranked pages are more likely to make use of image services which \rewriting opportunistically exploits. 

For the pages where \rewriting can be used, it offers significant data savings. For example, 5\% of pages in \aprFty and \topHun see savings of 20-30\% or 400KB saved per page, on average. With respect to user experience, images have a median SSIM value of~.97 (see Figure~\ref{fig:raw-savings-and-ssim-c}), a negligible quality reduction which is not detectable by a user. These high SSIM values are observed because these images are rewritten using standard means of compression.

Next, we quantify data savings and QoE impact of \fetching. While SSIM is useful to quantify visual impacts made to transformations that effect the \textit{entire} contents of an image (\eg blurring and color compression), it was not intended to reference quality of full images against partially complete images, as are produced by \fetching. We instead quantify visual impact for pages using the \textit{visual completeness}, a component of SpeedIndex~\cite{speed-index,netravali2018vesper} which is a Web performance metric describing the average time in which Web pages are rendered. Visual completeness is the comparison mechanism used by SpeedIndex to determine the fraction of a loading Web page that is rendered at a given point in time, allowing it to reasonably measure the visual impact of \textit{partially} complete images and pages. Under the hood, the visual completeness compares color histograms from screenshots of the Web page at points in its load to a screenshot of its fully rendered state. The visual completeness at a given time is the fraction of pixels with colors matching the final state. 

Figures~\ref{fig:range-request-results} (a) and ~\ref{fig:range-request-results} (b) compare the visual completeness values of pages with various amounts of image data requested, in 10\% increments, relative to the pages with~100\% of the image data requested. Each point on the CDF represents the visual completeness reached by~25\%,~50\%, and~75\% of pages. Figure~\ref{fig:range-request-results} (a) pictures all pages together, and (b) pages with a supermajority (>=60\%) of progressive images ($\approx$11\% of pages in our dataset). We can observe that across all pages, the median page is still 90\% visually complete with only~50\% of data requested. Likewise, pages in the~75th percentile are~90\% visually complete with only~30\% of the data requested. The median page with a supermajority of progressive images remains~90\% visually complete even with only~10--15\% of the image data requested. 

To expand on this result in terms of data saved, Figures~\ref{fig:range-request-results} (c) and (d) show CDFs of savings across pages assuming various levels of image data requested -- distinguishing between all pages (c) and the subset of pages hosting a supermajority of progressive images (d). Savings for general pages in (c) are shown for 25\%,~50\%, and~75\% of the image data requested. Given Figure~\ref{fig:range-request-results} has shown that, for progressive images, \fetching can be more aggressive and reach higher  visual completeness, savings in (d) are shown for~10\%,~25\%, and~50\% of image data fetched. We can observe that the median page sees a~28\% reduction in size by requesting~50\% of image data. This jumps to~42\% savings for~25\% of image data requested. The progressive pages saw ~$\approx$~40\% savings with~10\% data requested while remaining at least~90\% visually complete.

\subsection{User Experience}
\label{sec:results:user}
While visual completeness is a useful proxy for quantifying the impact \ourwork has on the user experience at scale, it is not a substitute for feedback from real users. Thus, we performed a user study to investigate how \ourwork affects the end user. 

We selected 40 pages with regular images and 10 pages with supermajority progressive images to show users in a crowdsourcing study, run via the Microworkers~\cite{microworkers} platform. Our study was a simple Web page that contained screenshots of the first viewport of the pages \textit{compressed} via \ourwork. For the screenshots, the \fetching component of \ourwork was configured such that 50\% of the image data was requested across all pages. The 40 regular pages were chosen randomly from~4 buckets of visual completeness to their \textit{original} counterparts. These buckets were chosen based on the distribution of pages in Figure~\ref{fig:range-request-results} (a), \ie VC >= 95\%, 90\% <= VC < 95\%, 85\% <= VC < 90\% and VC < 85\% for an average visual completeness of 89\%. The~10 progressive images had mean visual completeness of 97\%. On the study Web page, we showed the users the compressed version of each page with \ourwork side by side with the original page. The formal question we asked to users was, \textit{`How would you rate the quality of this compressed page which can extend your mobile data plan so you can browse more?'}. We provided the user a quality scale from~1--5 with meanings for each choice given in Figure~\ref{tab:user-study}.

For comparison, we also showed users screenshots of the same pages optimized with the Google Web Light tool~\cite{GoogleWebLight} (see Section~\ref{sec:related}). While Web Light acts as a good upper bound for what can truly be done in terms of data-savings, it is a less private approach, and not easily quantified in terms of \webcompat. We showed 41 total Web Light pages (by forcefully navigating pages through Web Lights servers at \url{http://googleweblight.com/i?u=URL}~\cite{Tahir2020-deconstructing, GoogleWebLight}, as 9 of the pages in our set of 50 could not be optimized by Web Light and simply redirected to the original page, a behavior documented by recent analyses~\cite{Tahir2020-deconstructing}. Not including such 9 pages, the Web Light pages had mean visual completeness of~73\%.  

Participants to our study were shown~20 of the~50 pages; for each of the 20, either our page or Web Light version (if available) was randomly shown, to avoid bias by having users compare two of the same URLs during the study. Two controls were also shown, one with a perfect rendition of a page (visual completeness of~100), and one with a page where all images were replaced by image placeholder icons (visual completeness of~29). We only accepted users who rated these as 4-5 and 1-2, respectively, which filtered out approximately~35\% of user responses to our study. We collected ratings from~200 Microworkers users (after filtering) giving each page approximately~40 ratings. We provide a link to an anonymous video of the study~\footnote{\url{https://streamable.com/e89yji}} which shows both pages optimized by \ourwork and by Web Light, as seen by our Microworker users.

Figure~\ref{tab:user-study} shows the median rating of each page calculated over the $\approx$40 user ratings collected per page. For \ourwork, we distinguish between the pages with supermajority progressive images and pages where the reflection trick was used (see Section~\ref{sec:impl:range}). From Figure~\ref{tab:user-study} we draw a few observations. The first is that users rated all 10 progressive pages very highly (mostly \textit{very good} and few \textit{good}), corroborating that progressive images can see higher data-savings for the same trade off in the user experience, discussed in Figure~\ref{fig:range-request-results}.  

The second is that the majority of pages with reflected images (21 out 40) were rated as \textit{usable} for most users. Though none of the pages were rated as \textit{broken}, about~20\% were given a \textit{poor} rating. Upon further investigation of these pages, we observed that this rating was typically given if a) human faces were distorted or b) actual text embedded within these images was reflected (and thus unreadable). Conversely, pages with text overlaid on the image, and not part of the image data where rated quite positively (4 or 5). While we do not take means to prevent such cases in \ourwork, we believe it to be a good direction for future work (see Section~\ref{sec:discussion}). Further, the only page that received a median score of 1 was our control page with broken placeholder images. This result suggests that reflections are generally more favorable for the user experience than placeholder images, a technique used by Chromium under its Data-Saving mode~images~\cite{blink-image-replacement}.

Finally, Figure~\ref{tab:user-study} shows that Web Light, while removing and rearranging much of the page contents and having a much lower visual completeness on average, was rated generally highly by users (21 good pages). However, as Google's servers have access to the contents before they are sent to the client, Web Light has more context, and time, to analyze the important parts of the page, factors not available to \ourwork.

\begin{figure}[b]
\centering
   \centering
    \resizebox{0.48\textwidth}{!}{\begin{tabular}{l c c c c c }
     \toprule
      \bf Method & \bf \begin{tabular}{@{}c@{}} Broken\\(1)\end{tabular}  & \bf \begin{tabular}{@{}c@{}} Poor \\(2)\end{tabular}   & \bf \begin{tabular}{@{}c@{}} Usable\\(3)\end{tabular}   & \bf\begin{tabular}{@{}c@{}} Good\\(4)\end{tabular}   & \bf \begin{tabular}{@{}c@{}} Very Good\\(5)\end{tabular}  \\
     \midrule
     \begin{tabular}{@{}c@{}} \ourwork\\(Reflections)\end{tabular}  & 0 & 8 & 21 & 8 & 3 \\
      \begin{tabular}{@{}c@{}} \ourwork\\(Progressive)\end{tabular}  & 0 & 0 & 0 & 3 & 7 \\
     Web Light & 0 & 2 & 11 & 21 & 7 \\
     \bottomrule
         \end{tabular} 
    }
    \vspace{1mm}
     \caption{User responses form our crowdsourced user study. The rows represent the type of optimization: \ourwork on pages with regular images, \ourwork on the progressive image pages, and Web Light optimized pages. }
    \label{tab:user-study}
\end{figure}

\subsection{Range Requests and Caching}
One caveat in applying range requests to save data is the potential impact on caching. For example, let's assume a user of \ourwork loads a page using a cellular connection, and then later loads the same page under a Wi-Fi connection, where no data-saving is necessary. Ideally, the portions of the images fetched using range requests are not re-fetched when opting out of \ourwork.

The heuristics that browsers employ for caching content ranges are not well documented. We thus set up an experiment on Chromium (version 83) to determine how caching rules for range requests are currently handled. First, we instrument Chromium to fetch real pages and images from our study under a cold cache. Next, we perform the following experiments:

\begin{enumerate}[leftmargin=0.5cm]
    \item We issued two successive range range requests for images with range of 0-10KB and 10-20KB and found they were both cached on subsequent requests, suggesting range requests of the same range are cacheable.
    \item We issued two range requests with overlapping ranges and found that the browser rewrites the range request to only fetch the remaining data, \eg a request for~0--10KB of an image following a range request for~0--20KB of the same was rewritten to request the range not yet in the cache:~10--20KB. 
    \item We issued a request for the full image following a range request and found that no data was wasted. For example first requesting~0--10KB of an image, followed by requesting the full image, only results in the remaining portion of the full image being requested by the browser.
    \item The reverse of the above is also true: we issued a request for the full image followed by a range request for~0--10KB of the same and found the range request to be served from the cache.
\end{enumerate}

These results add to the realism of a data-saving approach that leverages range requests, in that switching between \ourwork and normal browsing does not adversely affect HTTP caching, and hence waste data. This result further implies: a) flexible ranges can be used, say, if network conditions change, and b) \ourwork can be turned off, presumably by reading existing functionalities like the HTTP Save-Data header~\cite{simon-data-saving} with no excess data waste.

\subsection{Performance}

\begin{figure}[tb]
   \subfigure[]{\psfig{figure=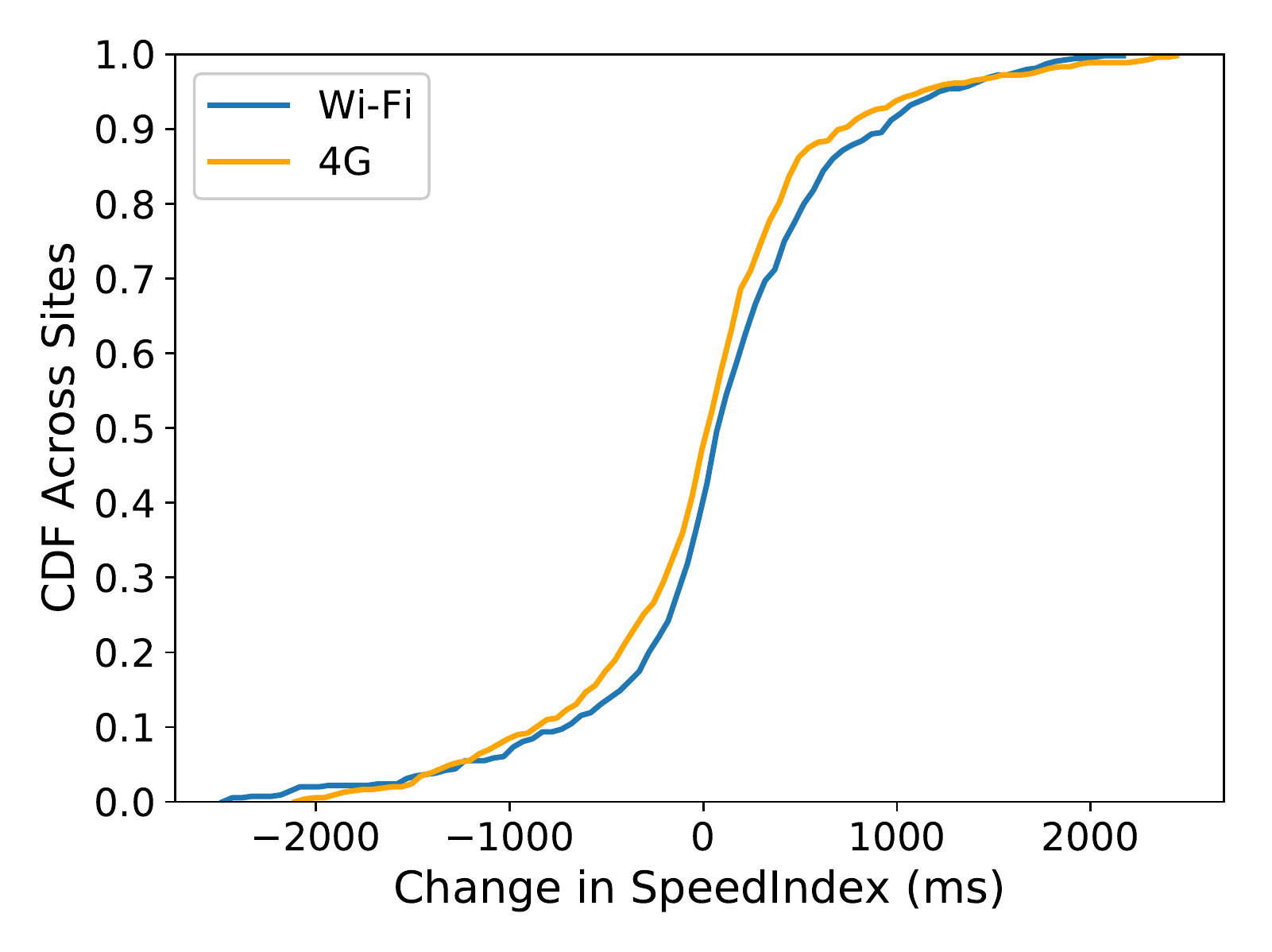, width=1.63in}\label{fig:system-overhead-a}}
   \subfigure[]{\psfig{figure=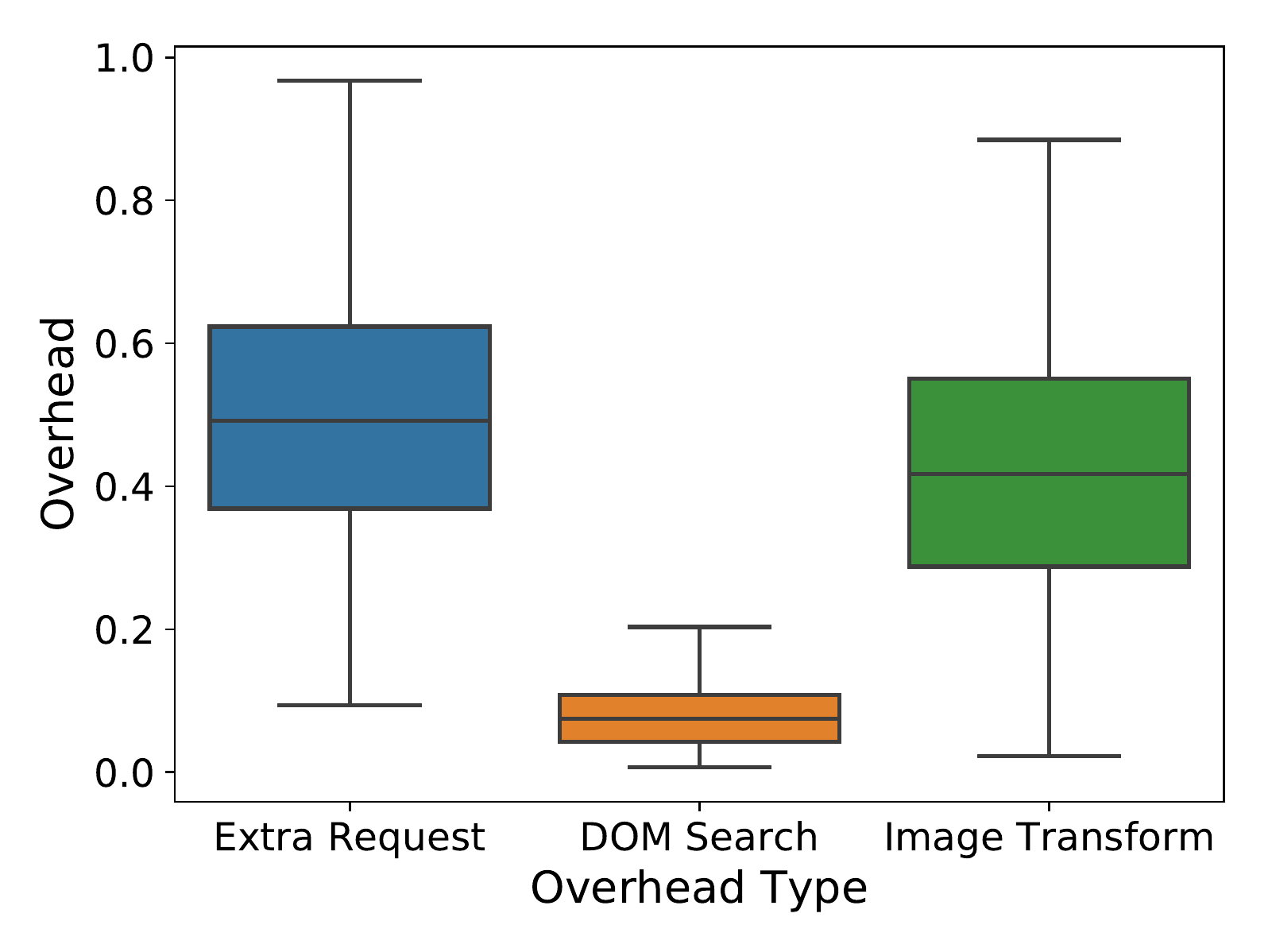, width=1.63in} \label{fig:system-overhead-b}}
   \caption{
   The CDF in (a) shows SpeedIndex is inflated by~70ms for the median page on Wi-Fi and~25ms on 4G. The boxplots in (b) show the relative contributions from each source of delay caused by \ourwork on the SpeedIndex.}
   \vspace{-0.05in}
   \label{fig:system-overhead}
   \vspace{-0.1in}
\end{figure}

\ourwork was designed with \textit{data-savings} rather \textit{speed} as its main goal. However, to be usable, it is still important that pages are not significantly slowed down, which we investigate here. 

\ourwork introduces a few extra operations to page loads which can potentially slow them down. The first is the extra range request used to fetch image metadata. Second is the search in the DOM to associate an image request to an object on the page. Third is the extra processing needed to create the reflections of non-progressive images. However, the data saved by \ourwork has the potential to offset such overhead. Note that while the extra request given an erroneous rule from \rewriting also adds some latency, this occurs relatively infrequently, as denoted by Figure~\ref{fig:rewrite-results}.

In order to quantify these additions to a Web page load, we measure the average overhead in rendering time of pages as given by the SpeedIndex~\cite{speed-index} metric. Figure~\ref{fig:system-overhead} shows the CDF of change in SpeedIndex when loading pages normally and with \ourwork across the $\sim$1,200 pages of our crawls. We compare across two network conditions, a Wi-Fi connection (\url{https://fast.com} reported 40Mbps up,~10Mbps down,~10ms RTT) and a Verizon 4G LTE connection (4Mbps up,~3Mbps down,~40ms RTT). We benchmark \ourwork with a visual completion budget of~90\%, implying that~50\% of the images are requested (see Figure~\ref{fig:range-request-results}).

\begin{figure}[t]
    \centering
    \includegraphics[width=2.3in]{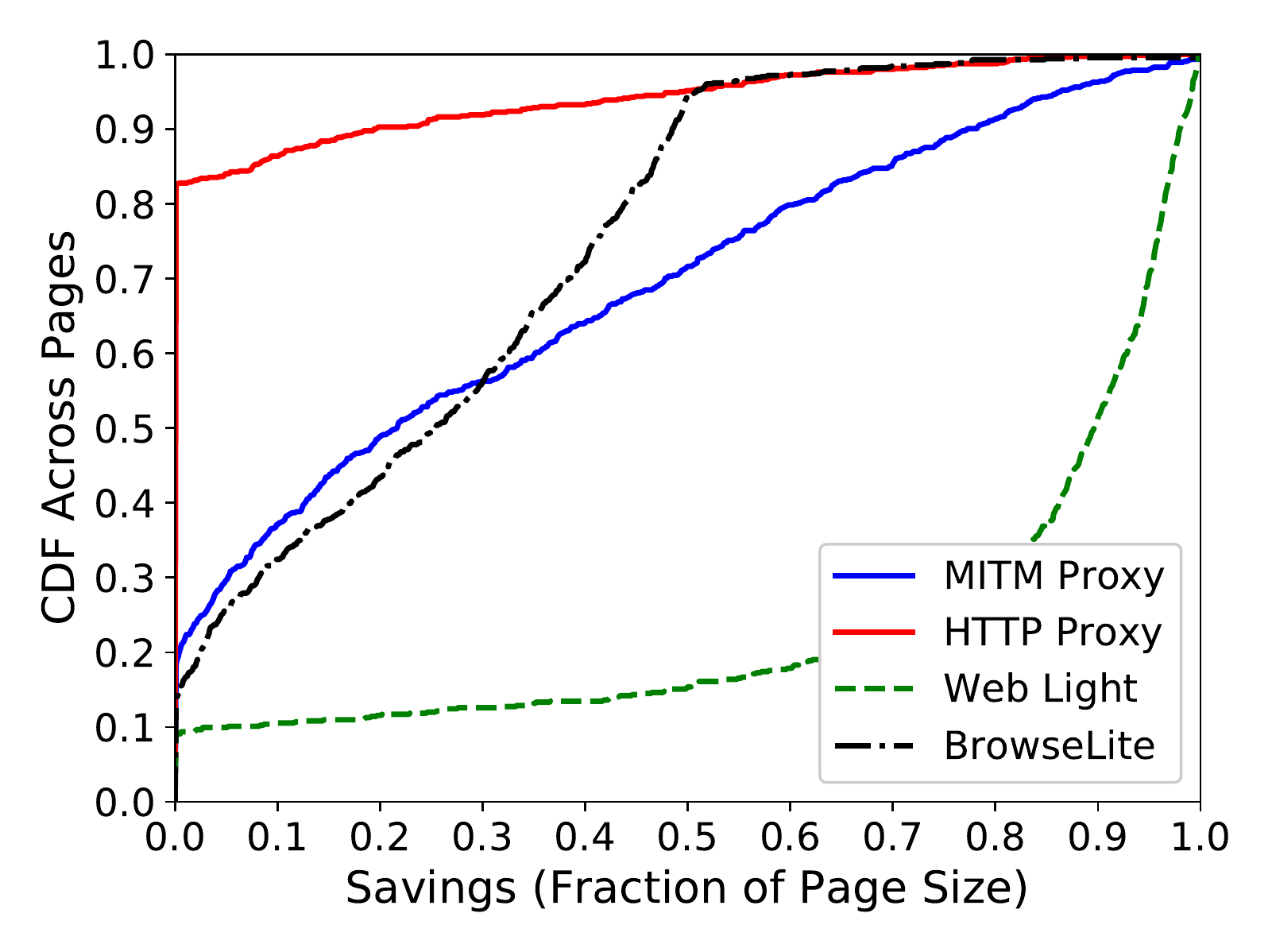}
    \caption{CDF of data-savings of middlebox approaches and Google Web Light compared to those of \ourwork.}
    \label{fig:savings-comparison}
\vspace{-0.1cm}
\end{figure}

We can observe that $\approx$80\% of pages on Wi-Fi experience overheads of <500 ms. Further, while 20\% of pages experience a more noticeable delay (>500 ms), 41-48\% of pages actually see an improved SpeedIndex by an average of $\approx$400ms. While 4G connections have higher RTT than Wi-Fi, implying the extra range requests of \ourwork are further delayed, they also have lower bandwidth, and thus benefit more from \ourwork's data savings. The result is that the distribution of change in SpeedIndex over WiFi and 4G are quantitatively similar. The improvement in SpeedIndex over both connections is due to the fact that some pages have images that actually render \textit{sooner} since only~50\% of the image content needs to be requested. While the fact that~20\% of pages experience noticeable slowdowns is significant, we note that the primary objective of \ourwork is to save bandwidth, and we expect users will tolerate a slight delay in their pages in exchange for data savings.

For the pages that see increased SpeedIndex, in Figure~\ref{fig:system-overhead} we analyze the 3 main causes of overhead from \ourwork for their relative attributions to the increases. From this, we can observe that the largest fraction of overhead is indeed due to the extra RTT (causing~50\% of the inflation in the median page), followed closely by transformation times using the browser's native image canvas APIs (45\% in the median page). The DOM search is relatively fast in comparison (5\% in the median page). The contributions to overhead were similar between the Wi-Fi and 4G experiments, with the extra range request taking up 5\% more overhead on 4G than on Wi-Fi. In section~\ref{sec:discussion} we provide a few potential ways the transformation and search overheads can be minimized going forward, mainly through tighter implementation of \ourwork in the browser.

\subsection{Comparisons}

Finally, we compare the data-savings of \ourwork\ -- the combined \fetching and \rewriting components~---~to the bandwidth savings made available through middlebox approaches as well as those offered by Google Web Light. For \ourwork, we target a visual completeness budget of at least 90\% and thus assume that 50\% of image data should be fetched. For middleboxes, we used the savings for the standard mode of our measurements Section~\ref{sec:measurement}. For Web Light pages, we navigated all the pages from our crawls through the Web Light system and derived savings as compared to the original versions of the same pages. All savings are in terms of relative page weights saved, including hot caches of inner-pages as described in Section~\ref{sec:measurement}.

Figure~\ref{fig:savings-comparison} shows the CDF of data-savings offered by \ourwork, middlebox approaches that act as (a) MITM proxies, and (b) HTTP only proxies, and Google Web Light, across the pages of our crawls. The figure shows that Web Light acts as an upper bound for potential savings on pages, with the median page seeing up to~90\% savings. As noted in Section~\ref{sec:related}, since all content is served through Web Light's servers, more resources (outside images), and even the actual style of the page, can be directly manipulated. This comes not only at the cost of privacy concerns, but also significant impact on \webcompat; from our experiments, pages served via Web Light achieve only 60\% visual completeness to their originals, on average. Web Light also failed to optimize $\approx$10\% of pages, as was observed for our user study and existing work~\cite{Tahir2020-deconstructing}. 

While middlebox approaches see about~4\% less savings compared to \ourwork at the median (21.6\% vs 25.4\%), the upper percentiles see up to~30\% more savings. While middlebox approaches are also limited to images, they intercept the content before reaching the client, allowing for more optimization opportunities at the cost of privacy concerns similar to those of Web Light. Further, if we look at only HTTP images from our dataset (we attempted HTTPS connections for all pages), the median savings of middlebox approaches drops to~0\%, and only~20\% at the~90th percentile, suggesting $\approx$62\% less pages are available to be optimized by not intercepting TLS. 

\section{Discussion}
\label{sec:discussion}
This section discusses some subtle privacy concerns of \ourwork as well as complexities of an in-browser implementation, along with some future work based on results from our user study.

\point{Privacy considerations}
While \ourwork is designed with privacy in mind, one subtle privacy concern lies in the caching of range requests. The current version of Chromium modifies range requests based on information from its cache in order to only fetch the next required portion of the range (see Section~\ref{sec:results}). Since resources in the HTTP cache can be hit across domains, this implies that a range request initiated on one domain can be resumed on another, thus leaking information on what other sites have been previously visited. 

This attack, known as a cross site leak or XS-Leak~\cite{ndss-2020-cosi}, is not specific to range requests. Many browsers have begun discussing the implementation of (or have already implemented in the case of Safari~\cite{implement-dual-key, davis-safari-caching}) dual-key caching. This policy prevents access to cross-origin resources from the HTTP cache, with main intent of stopping XS-Leaks. As this feature will also prevent such XS-leaks with range requests, we expect the \fetching feature of \ourwork to remain available and safe for the future. 

While \ourwork is implemented entirely client side for privacy, techniques for savings, such as \fetching and more curated rules for \rewriting, could be implemented privately at a CDN. However, the fact that 80\% of landing pages and 60\% of internal pages are not leveraging CDNs implies a client-side intervention is currently of import~\cite{http-archive-almanac}.

\point{Browser implementation} 
One concern for the adoption of \ourwork is the performance impact of >500ms for~20\% of pages (Figure~\ref{fig:system-overhead-a}). While the current version of \ourwork is implemented as a puppeteer application, a native implementation in the browser has the potential to eliminate overhead caused by image processing (for reflections) and DOM searches, which combined account for up to 50\% of overhead (see Figure~\ref{fig:system-overhead-b}). A native implementation can associate DOM elements directly with network requests, eliminating the need for a DOM search after the initial range request. Further, native use of image libraries (\eg libpng, libwebp) bundled with the browser will allow for faster reflections, compared with our current use of the high level canvas APIs and conversion of the images to dataURIs. If \ourwork were to see adoption, we believe these to be the next steps in its performance improvement. We do not perceive a way to avoid the extra range request to discover image metadata, which is a required step for \ourwork to work properly. 

\point{User Studies and Quality of Experience}
From our user study analyses, we made the observation that reflections were poorly rated when \fetching resulted in images with distorted faces or text. In the future, we wish to test these hypotheses with additional user studies. However, if true, what can be done to alleviate this impact on the user experience remains in question. One possibility we deem worth exploring is to use facial or textual detection models (\eg via CNNs~\cite{li2015convolutional}) on the partially downloaded image to identify presence of such features. Upon successful detection, an extra range request can be issued to better complete such images. Another approach is experimentation with context encoders~\cite{pathak2016context}, which can potentially complete our partially rendered images with no additional data cost. Either way, trade-offs in terms of the load times and bandwidth implications these proposed techniques may have on page loads will need to be explored.
\section{Conclusion}
\label{sec:conclusion}

Given the increased complexity and size of webpages, there has been an  enormous effort on the part of the web community to reduce data costs that the modern web now has on end users. However, existing approaches all trade off user privacy or web compatibility in exchange for such data savings. This paper presents \ourwork, a private data saving solution that focuses on optimization of \empty{images} during browsing. \ourwork reduces the data strain images place on the network by reducing their data requirements, through auto-configuring image services and by replacing standard HTTP requests for images with range requests. As shown through our experimentation, \ourwork is able to achieve~25\% data savings on the median webpage, with only a minor overhead on the page load time. Further, \ourwork is able to reduce data requirements, while keeping the median webpage \emph{usable}, as reported by real users. In future work, we plan to look at a tighter implementation of \ourwork in modern browsers and to explore the effects of a more advanced image processing pipeline to further reduce the potentially negative impacts on the end-user experience.

\bibliographystyle{ACM-Reference-Format}
\bibliography{main}


\begin{thebibliography}{60}


\ifx \showCODEN    \undefined \def \showCODEN     #1{\unskip}     \fi
\ifx \showDOI      \undefined \def \showDOI       #1{#1}\fi
\ifx \showISBNx    \undefined \def \showISBNx     #1{\unskip}     \fi
\ifx \showISBNxiii \undefined \def \showISBNxiii  #1{\unskip}     \fi
\ifx \showISSN     \undefined \def \showISSN      #1{\unskip}     \fi
\ifx \showLCCN     \undefined \def \showLCCN      #1{\unskip}     \fi
\ifx \shownote     \undefined \def \shownote      #1{#1}          \fi
\ifx \showarticletitle \undefined \def \showarticletitle #1{#1}   \fi
\ifx \showURL      \undefined \def \showURL       {\relax}        \fi
\providecommand\bibfield[2]{#2}
\providecommand\bibinfo[2]{#2}
\providecommand\natexlab[1]{#1}
\providecommand\showeprint[2][]{arXiv:#2}

\bibitem[\protect\citeauthoryear{??}{maj}{[n.d.]}]%
        {majestic_million}
 \bibinfo{year}{[n.d.]}\natexlab{}.
\newblock
  \bibinfo{howpublished}{\url{https://majestic.com/reports/majestic-million}}.
\newblock


\bibitem[\protect\citeauthoryear{??}{ale}{[n.d.]}]%
        {alexa}
 \bibinfo{year}{[n.d.]}\natexlab{}.
\newblock \bibinfo{howpublished}{\url{https://www.alexa.com/topsites}}.
\newblock


\bibitem[\protect\citeauthoryear{??}{lig}{[n.d.]}]%
        {ligthouse}
 \bibinfo{year}{[n.d.]}\natexlab{}.
\newblock
  \bibinfo{howpublished}{\url{https://developers.google.com/web/tools/lighthouse}}.
\newblock


\bibitem[\protect\citeauthoryear{??}{fas}{[n.d.]}]%
        {fastly}
 \bibinfo{year}{[n.d.]}\natexlab{}.
\newblock \bibinfo{howpublished}{\url{https://www.fastly.com/}}.
\newblock


\bibitem[\protect\citeauthoryear{??}{aka}{[n.d.]}]%
        {akamai}
 \bibinfo{year}{[n.d.]}\natexlab{}.
\newblock \bibinfo{howpublished}{\url{https://www.akamai.com/}}.
\newblock


\bibitem[\protect\citeauthoryear{??}{clo}{[n.d.]}]%
        {cloudflare}
 \bibinfo{year}{[n.d.]}\natexlab{}.
\newblock \bibinfo{howpublished}{\url{https://www.cloudflare.com//}}.
\newblock


\bibitem[\protect\citeauthoryear{??}{mic}{[n.d.]}]%
        {microworkers}
 \bibinfo{year}{[n.d.]}\natexlab{}.
\newblock \bibinfo{howpublished}{\url{https://microworkers.com/}}.
\newblock


\bibitem[\protect\citeauthoryear{??}{pup}{[n.d.]}]%
        {puppeteer}
 \bibinfo{year}{[n.d.]}\natexlab{}.
\newblock \bibinfo{title}{Puppeteer}.
\newblock \bibinfo{howpublished}{https://github.com/puppeteer/puppeteer}.
\newblock


\bibitem[\protect\citeauthoryear{??}{Chr}{2020}]%
        {ChromeLiteMode}
 \bibinfo{year}{2020}\natexlab{}.
\newblock \bibinfo{title}{Chrome LiteMode}.
\newblock \bibinfo{howpublished}{\url{tinyurl.com/1h8p6ykr }}.
\newblock


\bibitem[\protect\citeauthoryear{addy osmani}{addy osmani}{2020}]%
        {webdev-cls}
\bibfield{author}{\bibinfo{person}{addy osmani}.}
  \bibinfo{year}{2020}\natexlab{}.
\newblock \bibinfo{title}{Optimize Cumulative Layout Shift}.
\newblock \bibinfo{howpublished}{\url{tinyurl.com/2hcsh4zu }}.
\newblock


\bibitem[\protect\citeauthoryear{Agababov, Buettner, Chudnovsky, Cogan,
  Greenstein, McDaniel, Piatek, Scott, Welsh, and Yin}{Agababov
  et~al\mbox{.}}{2015}]%
        {Agababov2015-flywheel}
\bibfield{author}{\bibinfo{person}{Victor Agababov}, \bibinfo{person}{Michael
  Buettner}, \bibinfo{person}{Victor Chudnovsky}, \bibinfo{person}{Mark Cogan},
  \bibinfo{person}{Ben Greenstein}, \bibinfo{person}{Shane McDaniel},
  \bibinfo{person}{Michael Piatek}, \bibinfo{person}{Colin Scott},
  \bibinfo{person}{Matt Welsh}, {and} \bibinfo{person}{Bolian Yin}.}
  \bibinfo{year}{2015}\natexlab{}.
\newblock \showarticletitle{{Flywheel: Google's data compression proxy for the
  MobileWeb}}.
\newblock \bibinfo{journal}{\emph{Proceedings of the 12th USENIX Symposium on
  Networked Systems Design and Implementation, NSDI 2015}}
  (\bibinfo{year}{2015}).
\newblock


\bibitem[\protect\citeauthoryear{Agarwal}{Agarwal}{2015}]%
        {google-lite-break-pages}
\bibfield{author}{\bibinfo{person}{Ankur~P Agarwal}.}
  \bibinfo{year}{2015}\natexlab{}.
\newblock \bibinfo{title}{Google Web Lite - Move Fast, Break Things?}
\newblock
  \bibinfo{howpublished}{https://pricebaba.com/blog/google-weblight-move-fast-break-things}.
\newblock


\bibitem[\protect\citeauthoryear{Alakuijala and Rabaud}{Alakuijala and
  Rabaud}{2017}]%
        {google-webp-study}
\bibfield{author}{\bibinfo{person}{Jyrki Alakuijala} {and}
  \bibinfo{person}{Vincent Rabaud}.} \bibinfo{year}{2017}\natexlab{}.
\newblock \bibinfo{title}{Lossless and Transparency Encoding in WebP}.
\newblock
  \bibinfo{howpublished}{\url{https://developers.google.com/speed/webp/docs/webp_lossless_alpha_study}}.
\newblock


\bibitem[\protect\citeauthoryear{Aqeel, Chandrasekaran, Feldmann, and
  Maggs}{Aqeel et~al\mbox{.}}{2020}]%
        {ageel-imc-2020}
\bibfield{author}{\bibinfo{person}{Waqar Aqeel}, \bibinfo{person}{Balakrishnan
  Chandrasekaran}, \bibinfo{person}{Anja Feldmann}, {and}
  \bibinfo{person}{Bruce~M. Maggs}.} \bibinfo{year}{2020}\natexlab{}.
\newblock \showarticletitle{On Landing and Internal Web Pages: The Strange Case
  of Jekyll and Hyde in Web Performance Measurement}. In
  \bibinfo{booktitle}{\emph{Proceedings of the ACM Internet Measurement
  Conference}} \emph{(\bibinfo{series}{IMC '20})}.
  \bibinfo{publisher}{Association for Computing Machinery},
  \bibinfo{address}{New York, NY, USA}.
\newblock


\bibitem[\protect\citeauthoryear{Archive}{Archive}{2019}]%
        {http-archive-almanac}
\bibfield{author}{\bibinfo{person}{HTTP Archive}.}
  \bibinfo{year}{2019}\natexlab{}.
\newblock \bibinfo{title}{Web Almanac by HTTP Archive}.
\newblock \bibinfo{howpublished}{https://almanac.httparchive.org/en/2019/}.
\newblock


\bibitem[\protect\citeauthoryear{Arsenault}{Arsenault}{2017}]%
        {speed-index}
\bibfield{author}{\bibinfo{person}{Cody Arsenault}.}
  \bibinfo{year}{2017}\natexlab{}.
\newblock \bibinfo{title}{Speed Index Explained - Another Way to Measure Web
  Performance}.
\newblock \bibinfo{howpublished}{https://www.keycdn.com/blog/speed-index}.
\newblock


\bibitem[\protect\citeauthoryear{Atlas}{Atlas}{2019}]%
        {mobile-phone-stats}
\bibfield{author}{\bibinfo{person}{Device Atlas}.}
  \bibinfo{year}{2019}\natexlab{}.
\newblock \bibinfo{howpublished}{\url{https://tinyurl.com/59d5gy7b}}.
\newblock


\bibitem[\protect\citeauthoryear{Azad, Laperdrix, and Nikiforakis}{Azad
  et~al\mbox{.}}{2019}]%
        {azad-less-is-more}
\bibfield{author}{\bibinfo{person}{Babak~Amin Azad}, \bibinfo{person}{Pierre
  Laperdrix}, {and} \bibinfo{person}{Nick Nikiforakis}.}
  \bibinfo{year}{2019}\natexlab{}.
\newblock \showarticletitle{Less is More: Quantifying the Security Benefits of
  Debloating Web Applications}. In \bibinfo{booktitle}{\emph{28th {USENIX}
  Security Symposium ({USENIX} Security 19)}}. \bibinfo{publisher}{{USENIX}
  Association}, \bibinfo{address}{Santa Clara, CA}.
\newblock


\bibitem[\protect\citeauthoryear{Butkiewicz, Wang, Wu, Madhyastha, and
  Sekar}{Butkiewicz et~al\mbox{.}}{2015}]%
        {klotski-nsdi-2015}
\bibfield{author}{\bibinfo{person}{Michael Butkiewicz},
  \bibinfo{person}{Daimeng Wang}, \bibinfo{person}{Zhe Wu},
  \bibinfo{person}{Harsha~V. Madhyastha}, {and} \bibinfo{person}{Vyas Sekar}.}
  \bibinfo{year}{2015}\natexlab{}.
\newblock \showarticletitle{Klotski: Reprioritizing Web Content to Improve User
  Experience on Mobile Devices}. In \bibinfo{booktitle}{\emph{NSDI15}}.
  \bibinfo{publisher}{{USENIX} Association}, \bibinfo{address}{Oakland, CA}.
\newblock


\bibitem[\protect\citeauthoryear{Calvano}{Calvano}{2019}]%
        {http-archive-caching}
\bibfield{author}{\bibinfo{person}{Paul Calvano}.}
  \bibinfo{year}{2019}\natexlab{}.
\newblock \bibinfo{title}{Web Almanac by HTTP Archive: Part IV Chapter 16
  Caching}.
\newblock
  \bibinfo{howpublished}{https://almanac.httparchive.org/en/2019/caching}.
\newblock


\bibitem[\protect\citeauthoryear{Cardaci}{Cardaci}{2020}]%
        {cdp-support}
\bibfield{author}{\bibinfo{person}{Andrea Cardaci}.}
  \bibinfo{year}{2020}\natexlab{}.
\newblock \bibinfo{title}{Chrome-Remote-Interface}.
\newblock
  \bibinfo{howpublished}{\url{https://github.com/cyrus-and/chrome-remote-interface}}.
\newblock


\bibitem[\protect\citeauthoryear{Chaqfeh, Zaki, Hu, and Subramanian}{Chaqfeh
  et~al\mbox{.}}{2020}]%
        {Chaqfeh-2020-cleaner}
\bibfield{author}{\bibinfo{person}{Moumena Chaqfeh}, \bibinfo{person}{Yasir
  Zaki}, \bibinfo{person}{Jacinta Hu}, {and} \bibinfo{person}{Lakshmi
  Subramanian}.} \bibinfo{year}{2020}\natexlab{}.
\newblock \showarticletitle{{JSCleaner: De-Cluttering Mobile Webpages Through
  JavaScript Cleanup}}.
\newblock  (\bibinfo{year}{2020}).
\newblock


\bibitem[\protect\citeauthoryear{Chrome}{Chrome}{[n.d.]}]%
        {GoogleWebLight}
\bibfield{author}{\bibinfo{person}{Google Chrome}.}
  \bibinfo{year}{[n.d.]}\natexlab{}.
\newblock \bibinfo{title}{Web Light: Faster and lighter mobile pages from
  search}.
\newblock
  \bibinfo{howpublished}{https://support.google.com/webmasters/answer/6211428}.
\newblock


\bibitem[\protect\citeauthoryear{Coyer}{Coyer}{2020}]%
        {data-uris}
\bibfield{author}{\bibinfo{person}{Chris Coyer}.}
  \bibinfo{year}{2020}\natexlab{}.
\newblock \bibinfo{title}{Data URIs}.
\newblock \bibinfo{howpublished}{\url{https://css-tricks.com/data-uris/}}.
\newblock


\bibitem[\protect\citeauthoryear{Coyier}{Coyier}{2017}]%
        {css-sprites}
\bibfield{author}{\bibinfo{person}{Chris Coyier}.}
  \bibinfo{year}{2017}\natexlab{}.
\newblock \bibinfo{title}{CSS Sprites: What They Are, Why They’re Cool, and
  How To Use Them}.
\newblock \bibinfo{howpublished}{\url{https://css-tricks.com/css-sprites/}}.
\newblock


\bibitem[\protect\citeauthoryear{Davies}{Davies}{2018}]%
        {davis-safari-caching}
\bibfield{author}{\bibinfo{person}{Andy Davies}.}
  \bibinfo{year}{2018}\natexlab{}.
\newblock \bibinfo{title}{Safari\, Caching\, and Third-Party Resources}.
\newblock \bibinfo{howpublished}{https://bit.ly/3cLJW22/}.
\newblock


\bibitem[\protect\citeauthoryear{Erdmann}{Erdmann}{2019}]%
        {smashing-image-previews}
\bibfield{author}{\bibinfo{person}{Christoph Erdmann}.}
  \bibinfo{year}{2019}\natexlab{}.
\newblock \bibinfo{title}{Faster Image Loading with Embedded Image Previews}.
\newblock \bibinfo{howpublished}{\url{https://bit.ly/3kfeNoz}}.
\newblock


\bibitem[\protect\citeauthoryear{Felt, Barnes, King, Palmer, Bentzel, and
  Tabriz}{Felt et~al\mbox{.}}{2017}]%
        {felt-https-2016}
\bibfield{author}{\bibinfo{person}{Adrienne~Porter Felt},
  \bibinfo{person}{Richard Barnes}, \bibinfo{person}{April King},
  \bibinfo{person}{Chris Palmer}, \bibinfo{person}{Chris Bentzel}, {and}
  \bibinfo{person}{Parisa Tabriz}.} \bibinfo{year}{2017}\natexlab{}.
\newblock \showarticletitle{Measuring {HTTPS} Adoption on the Web}. In
  \bibinfo{booktitle}{\emph{USENIX Security 17}}. \bibinfo{publisher}{{USENIX}
  Association}.
\newblock


\bibitem[\protect\citeauthoryear{Goel and Steiner}{Goel and Steiner}{2020}]%
        {Goel2020}
\bibfield{author}{\bibinfo{person}{Utkarsh Goel} {and} \bibinfo{person}{Moritz
  Steiner}.} \bibinfo{year}{2020}\natexlab{}.
\newblock \showarticletitle{{System to Identify and Elide Superfluous
  JavaScript Code for Faster Webpage Loads}}.
\newblock  (\bibinfo{date}{mar} \bibinfo{year}{2020}).
\newblock


\bibitem[\protect\citeauthoryear{Groups}{Groups}{2019}]%
        {implement-dual-key}
\bibfield{author}{\bibinfo{person}{Google Groups}.}
  \bibinfo{year}{2019}\natexlab{}.
\newblock \bibinfo{title}{Intent to Implement Double-Keyed HTTP Cache}.
\newblock \bibinfo{howpublished}{\url{https://rb.gy/5ngyus}}.
\newblock


\bibitem[\protect\citeauthoryear{Hearne}{Hearne}{2020}]%
        {simon-data-saving}
\bibfield{author}{\bibinfo{person}{Simon Hearne}.}
  \bibinfo{year}{2020}\natexlab{}.
\newblock \bibinfo{title}{Who Opts-in to Save-Data?}
\newblock \bibinfo{howpublished}{tinyurl.com/10bxrn7d}.
\newblock


\bibitem[\protect\citeauthoryear{Hempenius}{Hempenius}{2019}]%
        {image-cdns}
\bibfield{author}{\bibinfo{person}{Katie Hempenius}.}
  \bibinfo{year}{2019}\natexlab{}.
\newblock \bibinfo{title}{CDNs for Optimizing Images}.
\newblock \bibinfo{howpublished}{\url{tinyurl.com/4n9g6qhp }}.
\newblock


\bibitem[\protect\citeauthoryear{Kadlec}{Kadlec}{2019}]%
        {tim-kadlec-making-sense}
\bibfield{author}{\bibinfo{person}{Tim Kadlec}.}
  \bibinfo{year}{2019}\natexlab{}.
\newblock \bibinfo{title}{Making Sense of Chrome Lite Pages}.
\newblock \bibinfo{howpublished}{tinyurl.com/26hzk53b}.
\newblock


\bibitem[\protect\citeauthoryear{Kadlec}{Kadlec}{2020}]%
        {what-does-my-site-cost}
\bibfield{author}{\bibinfo{person}{Tim Kadlec}.}
  \bibinfo{year}{2020}\natexlab{}.
\newblock \bibinfo{title}{What Does My Site Cost}.
\newblock \bibinfo{howpublished}{https://whatdoesmysitecost.com/}.
\newblock


\bibitem[\protect\citeauthoryear{Kandrot}{Kandrot}{2015}]%
        {facebook-previews}
\bibfield{author}{\bibinfo{person}{Edward Kandrot}.}
  \bibinfo{year}{2015}\natexlab{}.
\newblock \bibinfo{title}{The technology behind preview photos}.
\newblock \bibinfo{howpublished}{\url{https://bit.ly/37lG9FV}}.
\newblock


\bibitem[\protect\citeauthoryear{Kelton, Ryoo, Balasubramanian, and Das}{Kelton
  et~al\mbox{.}}{2017}]%
        {kelton-webgaze-2017}
\bibfield{author}{\bibinfo{person}{Conor Kelton}, \bibinfo{person}{Jihoon
  Ryoo}, \bibinfo{person}{Aruna Balasubramanian}, {and}
  \bibinfo{person}{Samir~R. Das}.} \bibinfo{year}{2017}\natexlab{}.
\newblock \showarticletitle{Improving User Perceived Page Load Times Using
  Gaze}. In \bibinfo{booktitle}{\emph{14th {USENIX} Symposium on Networked
  Systems Design and Implementation ({NSDI} 17)}}. \bibinfo{publisher}{{USENIX}
  Association}, \bibinfo{address}{Boston, MA}.
\newblock


\bibitem[\protect\citeauthoryear{Kondracki, Aliyeva, Egele, Polakis, and
  Nikiforakis}{Kondracki et~al\mbox{.}}{2020}]%
        {Kondracki-2020-meddling}
\bibfield{author}{\bibinfo{person}{B. Kondracki}, \bibinfo{person}{A. Aliyeva},
  \bibinfo{person}{M. Egele}, \bibinfo{person}{J. Polakis}, {and}
  \bibinfo{person}{N. Nikiforakis}.} \bibinfo{year}{2020}\natexlab{}.
\newblock \showarticletitle{Meddling Middlemen: Empirical Analysis of the Risks
  of Data-Saving Mobile Browsers}. In \bibinfo{booktitle}{\emph{2020 IEEE
  Symposium on Security and Privacy (SP)}}. \bibinfo{publisher}{IEEE Computer
  Society}, \bibinfo{address}{Los Alamitos, CA, USA}.
\newblock


\bibitem[\protect\citeauthoryear{Li, Lin, Shen, Brandt, and Hua}{Li
  et~al\mbox{.}}{2015}]%
        {li2015convolutional}
\bibfield{author}{\bibinfo{person}{Haoxiang Li}, \bibinfo{person}{Zhe Lin},
  \bibinfo{person}{Xiaohui Shen}, \bibinfo{person}{Jonathan Brandt}, {and}
  \bibinfo{person}{Gang Hua}.} \bibinfo{year}{2015}\natexlab{}.
\newblock \showarticletitle{A convolutional neural network cascade for face
  detection}. In \bibinfo{booktitle}{\emph{Proceedings of the IEEE conference
  on computer vision and pattern recognition}}.
\newblock


\bibitem[\protect\citeauthoryear{Li, Wang, Xu, Zhong, Li, Liu, Wilson, and
  Zhao}{Li et~al\mbox{.}}{2016}]%
        {li-nsdi-2016}
\bibfield{author}{\bibinfo{person}{Zhenhua Li}, \bibinfo{person}{Weiwei Wang},
  \bibinfo{person}{Tianyin Xu}, \bibinfo{person}{Xin Zhong},
  \bibinfo{person}{Xiang-Yang Li}, \bibinfo{person}{Yunhao Liu},
  \bibinfo{person}{Christo Wilson}, {and} \bibinfo{person}{Ben~Y. Zhao}.}
  \bibinfo{year}{2016}\natexlab{}.
\newblock \showarticletitle{Exploring Cross-Application Cellular Traffic
  Optimization with Baidu TrafficGuard}. In \bibinfo{booktitle}{\emph{13th
  {USENIX} Symposium on Networked Systems Design and Implementation ({NSDI}
  16)}}. \bibinfo{publisher}{{USENIX} Association}, \bibinfo{address}{Santa
  Clara, CA}.
\newblock


\bibitem[\protect\citeauthoryear{Little}{Little}{2016}]%
        {blink-image-replacement}
\bibfield{author}{\bibinfo{person}{Scott Little}.}
  \bibinfo{year}{2016}\natexlab{}.
\newblock \bibinfo{title}{Image Replacement in Blink}.
\newblock \bibinfo{howpublished}{https://rb.gy/lj4cdm}.
\newblock


\bibitem[\protect\citeauthoryear{Mike~Taylor}{Mike~Taylor}{2014}]%
        {mozilla-webcompat}
\bibfield{author}{\bibinfo{person}{Robert~Nyman Mike~Taylor}.}
  \bibinfo{year}{2014}\natexlab{}.
\newblock \bibinfo{title}{Intoducing webcompat.com}.
\newblock
  \bibinfo{howpublished}{https://hacks.mozilla.org/2014/06/introducing-webcompat-com/}.
\newblock


\bibitem[\protect\citeauthoryear{Netravali, Goyal, Mickens, and
  Balakrishnan}{Netravali et~al\mbox{.}}{[n.d.]}]%
        {Netravali-polaris}
\bibfield{author}{\bibinfo{person}{Ravi Netravali}, \bibinfo{person}{Ameesh
  Goyal}, \bibinfo{person}{James Mickens}, {and} \bibinfo{person}{Hari
  Balakrishnan}.} \bibinfo{year}{[n.d.]}\natexlab{}.
\newblock \showarticletitle{Polaris: Faster Page Loads Using Fine-grained
  Dependency Tracking}. In \bibinfo{booktitle}{\emph{NSDI16}}.
\newblock


\bibitem[\protect\citeauthoryear{Netravali and Mickens}{Netravali and
  Mickens}{2018}]%
        {Netravali2018-rc2}
\bibfield{author}{\bibinfo{person}{Ravi Netravali} {and} \bibinfo{person}{James
  Mickens}.} \bibinfo{year}{2018}\natexlab{}.
\newblock \showarticletitle{{Remote-control caching: Proxy-based url rewriting
  to decrease mobile browsing bandwidth}}. In
  \bibinfo{booktitle}{\emph{HotMobile 2018 - Proceedings of the 19th
  International Workshop on Mobile Computing Systems and Applications}},
  Vol.~\bibinfo{volume}{2018-February}. \bibinfo{publisher}{Association for
  Computing Machinery, Inc}.
\newblock


\bibitem[\protect\citeauthoryear{Netravali and Mickens}{Netravali and
  Mickens}{2019}]%
        {Netravali2019-prophecy}
\bibfield{author}{\bibinfo{person}{Ravi Netravali} {and} \bibinfo{person}{James
  Mickens}.} \bibinfo{year}{2019}\natexlab{}.
\newblock \showarticletitle{{Prophecy: Accelerating mobile page loads using
  final-state write logs}}.
\newblock \bibinfo{journal}{\emph{Proceedings of the 15th USENIX Symposium on
  Networked Systems Design and Implementation, NSDI 2018}}
  (\bibinfo{year}{2019}).
\newblock


\bibitem[\protect\citeauthoryear{Netravali, Nathan, Mickens, and
  Balakrishnan}{Netravali et~al\mbox{.}}{2018}]%
        {netravali2018vesper}
\bibfield{author}{\bibinfo{person}{Ravi Netravali}, \bibinfo{person}{Vikram
  Nathan}, \bibinfo{person}{James Mickens}, {and} \bibinfo{person}{Hari
  Balakrishnan}.} \bibinfo{year}{2018}\natexlab{}.
\newblock \showarticletitle{Vesper: Measuring time-to-interactivity for web
  pages}. In \bibinfo{booktitle}{\emph{15th $\{$USENIX$\}$ Symposium on
  Networked Systems Design and Implementation ($\{$NSDI$\}$ 18)}}.
\newblock


\bibitem[\protect\citeauthoryear{Opera}{Opera}{[n.d.]}]%
        {opera-turbo}
\bibfield{author}{\bibinfo{person}{Opera}.} \bibinfo{year}{[n.d.]}\natexlab{}.
\newblock \bibinfo{title}{Data savings and turbo mode}.
\newblock \bibinfo{howpublished}{https://www.opera.com/turbo}.
\newblock


\bibitem[\protect\citeauthoryear{Pathak, Krahenbuhl, Donahue, Darrell, and
  Efros}{Pathak et~al\mbox{.}}{2016}]%
        {pathak2016context}
\bibfield{author}{\bibinfo{person}{Deepak Pathak}, \bibinfo{person}{Philipp
  Krahenbuhl}, \bibinfo{person}{Jeff Donahue}, \bibinfo{person}{Trevor
  Darrell}, {and} \bibinfo{person}{Alexei~A Efros}.}
  \bibinfo{year}{2016}\natexlab{}.
\newblock \showarticletitle{Context encoders: Feature learning by inpainting}.
  In \bibinfo{booktitle}{\emph{Proceedings of the IEEE conference on computer
  vision and pattern recognition}}.
\newblock


\bibitem[\protect\citeauthoryear{Perez}{Perez}{2015}]%
        {medium-image-previews}
\bibfield{author}{\bibinfo{person}{Jose~M. Perez}.}
  \bibinfo{year}{2015}\natexlab{}.
\newblock \bibinfo{title}{Medium progressive image loading}.
\newblock \bibinfo{howpublished}{\url{tinyurl.com/1l1ht86v }}.
\newblock


\bibitem[\protect\citeauthoryear{Ruamviboonsuk, Netravali, Uluyol, and
  Madhyastha}{Ruamviboonsuk et~al\mbox{.}}{2017}]%
        {vroom-sigcomm-17}
\bibfield{author}{\bibinfo{person}{Vaspol Ruamviboonsuk}, \bibinfo{person}{Ravi
  Netravali}, \bibinfo{person}{Muhammed Uluyol}, {and}
  \bibinfo{person}{Harsha~V. Madhyastha}.} \bibinfo{year}{2017}\natexlab{}.
\newblock \showarticletitle{Vroom: Accelerating the Mobile Web with
  Server-Aided Dependency Resolution}. In \bibinfo{booktitle}{\emph{Proceedings
  of the Conference of the ACM Special Interest Group on Data Communication}}
  (Los Angeles, CA, USA) \emph{(\bibinfo{series}{SIGCOMM ’17})}.
  \bibinfo{publisher}{Association for Computing Machinery},
  \bibinfo{address}{New York, NY, USA}.
\newblock
\showISBNx{9781450346535}


\bibitem[\protect\citeauthoryear{Scheitle, Hohlfeld, Gamba, Jelten, Zimmermann,
  Strowes, and Vallina{-}Rodriguez}{Scheitle et~al\mbox{.}}{2018}]%
        {imc2018toplists}
\bibfield{author}{\bibinfo{person}{Quirin Scheitle}, \bibinfo{person}{Oliver
  Hohlfeld}, \bibinfo{person}{Julien Gamba}, \bibinfo{person}{Jonas Jelten},
  \bibinfo{person}{Torsten Zimmermann}, \bibinfo{person}{Stephen~D. Strowes},
  {and} \bibinfo{person}{Narseo Vallina{-}Rodriguez}.}
  \bibinfo{year}{2018}\natexlab{}.
\newblock \showarticletitle{A Long Way to the Top: Significance, Structure, and
  Stability of Internet Top Lists}.
\newblock \bibinfo{journal}{\emph{CoRR}}  \bibinfo{volume}{abs/1805.11506}
  (\bibinfo{year}{2018}).
\newblock


\bibitem[\protect\citeauthoryear{Singh, Madhyastha, Krishnamurthy, and
  Govindan}{Singh et~al\mbox{.}}{2015}]%
        {Singh2015-flexiweb}
\bibfield{author}{\bibinfo{person}{Shailendra Singh},
  \bibinfo{person}{Harsha~V. Madhyastha}, \bibinfo{person}{Srikanth~V.
  Krishnamurthy}, {and} \bibinfo{person}{Ramesh Govindan}.}
  \bibinfo{year}{2015}\natexlab{}.
\newblock \showarticletitle{{Flexiweb: Network-aware compaction for
  accelerating mobile web transfers}}. In \bibinfo{booktitle}{\emph{Proceedings
  of the Annual International Conference on Mobile Computing and Networking,
  MOBICOM}}, Vol.~\bibinfo{volume}{2015-September}.
  \bibinfo{publisher}{Association for Computing Machinery}.
\newblock
\showISBNx{9781450336192}


\bibitem[\protect\citeauthoryear{Sudhodanan, Khodayari, and
  Caballero}{Sudhodanan et~al\mbox{.}}{2019}]%
        {ndss-2020-cosi}
\bibfield{author}{\bibinfo{person}{Avinash Sudhodanan}, \bibinfo{person}{Soheil
  Khodayari}, {and} \bibinfo{person}{Juan Caballero}.}
  \bibinfo{year}{2019}\natexlab{}.
\newblock \showarticletitle{Cross-Origin State Inference (COSI) Attacks:
  Leaking Web Site States through XS-Leaks}.
\newblock \bibinfo{journal}{\emph{arXiv preprint arXiv:1908.02204}}
  (\bibinfo{year}{2019}).
\newblock


\bibitem[\protect\citeauthoryear{Sy, Burkert, Federrath, and Fischer}{Sy
  et~al\mbox{.}}{2018}]%
        {sy-tls-2018}
\bibfield{author}{\bibinfo{person}{Erik Sy}, \bibinfo{person}{Christian
  Burkert}, \bibinfo{person}{Hannes Federrath}, {and} \bibinfo{person}{Mathias
  Fischer}.} \bibinfo{year}{2018}\natexlab{}.
\newblock \showarticletitle{Tracking Users across the Web via {TLS} Session
  Resumption}.
\newblock \bibinfo{journal}{\emph{CoRR}}  \bibinfo{volume}{abs/1810.07304}
  (\bibinfo{year}{2018}).
\newblock


\bibitem[\protect\citeauthoryear{Tahir, Munir, Malik, Qazi, and Qazi}{Tahir
  et~al\mbox{.}}{2020}]%
        {Tahir2020-deconstructing}
\bibfield{author}{\bibinfo{person}{Ammar Tahir},
  \bibinfo{person}{Muhammad~Tahir Munir}, \bibinfo{person}{Shaiq~Munir Malik},
  \bibinfo{person}{Zafar~Ayyub Qazi}, {and} \bibinfo{person}{Ihsan~Ayyub
  Qazi}.} \bibinfo{year}{2020}\natexlab{}.
\newblock \showarticletitle{{Deconstructing Google's Web Light Service}}.
\newblock  (\bibinfo{year}{2020}).
\newblock


\bibitem[\protect\citeauthoryear{Umawing}{Umawing}{2017}]%
        {http-not-secure}
\bibfield{author}{\bibinfo{person}{Jovi Umawing}.}
  \bibinfo{year}{2017}\natexlab{}.
\newblock \bibinfo{title}{Google Reminds Website Owners to Switch to HTTPS
  Before October Deadline}.
\newblock \bibinfo{howpublished}{https://rb.gy/saxnj5}.
\newblock


\bibitem[\protect\citeauthoryear{Urban, Degeling, Holz, and Pohlmann}{Urban
  et~al\mbox{.}}{2020}]%
        {urban2020beyond}
\bibfield{author}{\bibinfo{person}{Tobias Urban}, \bibinfo{person}{Martin
  Degeling}, \bibinfo{person}{Thorsten Holz}, {and} \bibinfo{person}{Norbert
  Pohlmann}.} \bibinfo{year}{2020}\natexlab{}.
\newblock \showarticletitle{Beyond the front page: Measuring third party
  dynamics in the field}. In \bibinfo{booktitle}{\emph{Proceedings of The Web
  Conference 2020}}.
\newblock


\bibitem[\protect\citeauthoryear{Wang, Balasubramanian, Krishnamurthy, and
  Wetherall}{Wang et~al\mbox{.}}{2013}]%
        {wprof-nsdi-2013}
\bibfield{author}{\bibinfo{person}{Xiao~Sophia Wang}, \bibinfo{person}{Aruna
  Balasubramanian}, \bibinfo{person}{Arvind Krishnamurthy}, {and}
  \bibinfo{person}{David Wetherall}.} \bibinfo{year}{2013}\natexlab{}.
\newblock \showarticletitle{Demystifying Page Load Performance with WProf}. In
  \bibinfo{booktitle}{\emph{10th {USENIX} Symposium on Networked Systems Design
  and Implementation ({NSDI} 13)}}. \bibinfo{publisher}{{USENIX} Association},
  \bibinfo{address}{Lombard, IL}.
\newblock


\bibitem[\protect\citeauthoryear{Wang, Krishnamurthy, and Wetherall}{Wang
  et~al\mbox{.}}{2016}]%
        {wang-shandian-2016}
\bibfield{author}{\bibinfo{person}{Xiao~Sophia Wang}, \bibinfo{person}{Arvind
  Krishnamurthy}, {and} \bibinfo{person}{David Wetherall}.}
  \bibinfo{year}{2016}\natexlab{}.
\newblock \showarticletitle{Speeding up Web Page Loads with Shandian}. In
  \bibinfo{booktitle}{\emph{NSDI16}}. \bibinfo{address}{Santa Clara, CA}.
\newblock


\bibitem[\protect\citeauthoryear{Wang, Lin, Zhong, and Chishtie}{Wang
  et~al\mbox{.}}{2012}]%
        {www-12-how-far}
\bibfield{author}{\bibinfo{person}{Zhen Wang}, \bibinfo{person}{Felix~Xiaozhu
  Lin}, \bibinfo{person}{Lin Zhong}, {and} \bibinfo{person}{Mansoor Chishtie}.}
  \bibinfo{year}{2012}\natexlab{}.
\newblock \showarticletitle{How Far Can Client-Only Solutions Go for Mobile
  Browser Speed?}. In \bibinfo{booktitle}{\emph{Proceedings of the 21st
  International Conference on World Wide Web}} (Lyon, France)
  \emph{(\bibinfo{series}{WWW ’12})}. \bibinfo{publisher}{Association for
  Computing Machinery}, \bibinfo{address}{New York, NY, USA}.
\newblock
\showISBNx{9781450312295}


\bibitem[\protect\citeauthoryear{{Zhou Wang}, {Bovik}, {Sheikh}, and
  {Simoncelli}}{{Zhou Wang} et~al\mbox{.}}{2004}]%
        {ssim-metric}
\bibfield{author}{\bibinfo{person}{{Zhou Wang}}, \bibinfo{person}{A.~C.
  {Bovik}}, \bibinfo{person}{H.~R. {Sheikh}}, {and} \bibinfo{person}{E.~P.
  {Simoncelli}}.} \bibinfo{year}{2004}\natexlab{}.
\newblock \showarticletitle{Image quality assessment: from error visibility to
  structural similarity}.
\newblock \bibinfo{journal}{\emph{IEEE Transactions on Image Processing}}
  \bibinfo{volume}{13} (\bibinfo{year}{2004}).
\newblock


\end{thebibliography}

\end{document}